\documentclass[pre,preprint,a4paper,showpacs,showkeys,superscriptaddress]{revtex4-1}
\usepackage{epsfig}
\usepackage{amsmath}
\usepackage{amssymb}
\begin{document}
\title{Chaos in Kicked Ratchets}
\author{D.G. Zarlenga}
\author{H.A. Larrondo}
\thanks{CONICET Researcher}
\email{larrondo@fi.mdp.edu.ar}
\author{C.M. Arizmendi}
\affiliation{Departamento de F\'{\i}sica e Instituto de Investigaciones Cient\'ificas y Tecnol\'ogicas en  Electr\'onica, Facultad de
Ingenier\'{\i}a, Universidad Nacional de Mar del Plata,\\  Av. J.B.
Justo 4302, 7600 Mar del Plata, Argentina\\}
\author{Fereydoon Family}
\affiliation{Department of Physics, Emory University, Atlanta, GA
30322,  USA}
\date{\today}
\begin{abstract}
We present a {\it minimal} one-dimensional deterministic continuous dynamical system that exhibits chaotic behavior and complex transport properties. Our model is an overdamped rocking ratchet that is periodically kicked with a delta function potential. We develop an analytical approach that predicts many key features of the system, such as current reversals, as well as the presence of chaotic behavior and bifurcation. We show that our approach can be easily extended to other types of periodic forces, including the square wave.
\end{abstract}
\date{\today}
%
\maketitle
\section{Introduction}
\label{sec:intro}
The nonequilibrium mechanism of generating directed transport from
the interaction of broken symmetry, periodic structures, and fluctuations in the presence of an unbiased driving force, usually known as the ratchet effect,  has recently received much
attention \cite{Reimann2002,Astumian1997b,Astumian2002,Hanggi2005, Zarlenga2009}. 
This growing interest in ratchets is mostly due
to the large number of successful applications of ratchet models to understand and control a wide variety of physical and biological systems.  For example, ratchets have been used to model molecular or Brownian
motors  inside the eukaryotic cells \cite{Astumian1997b,Astumian2002,Nishinari2005,Greulich2007,Sparacino2011},
as well as the operation of muscles  at the body level \cite{Hanggi1996}. Another important application is on the development of devices for guiding nano/micro particles, such as  transport of cold atoms
in optical lattices \cite{Cohen2004,Carlo2005}, control of the motion of vortices in superconducting devices \cite{Perezdelara2011,Baert1995,Martin1997,Avci2010,Reichhardt1998,Berdiyorov2006,Baert1995b,Reichhardt2001} and mass separation and trapping schemes at the microscale \cite{Hanggi2005, Gorre1997,Derenyi1998,Ertas1998,Duke1998}.
Thermal fluctuations produce a directed current in the motion of Brownian particles when the thermal noise interacts with the ratchet potential. On the other hand, even in the absence of noise, underdamped or inertial ratchets show complex dynamical behavior, including chaotic motion \cite{Jung1996}. This deterministically induced chaos
to some extent replaces the role of
noise and produces certain unusual types of dynamical behavior, including multiple current reversals  \cite{Mateos2000,Barbi2000}, which is particularly useful for technological applications such as biological particle separation \cite{Gorre1997,Derenyi1998,Ertas1998,Duke1998}.
Recently, Vincent et al. \cite{Vincent2010} considered a system of
two interacting inertial ratchets, and demonstrated how the coupling can be used to control  current reversals.

Chaotic behavior  in continuous dynamical systems is observed if the system possesses a certain minimum degree of nonlinearity. The reason is that the Poincar\'e-Bendixson theorem stipulates that chaotic behavior does not exist in one or two-dimensional continuous dynamical systems, because such systems have regular solutions.   This is in contrast to discrete systems, such as the logistic map, that show chaotic behavior regardless of their dimensionality. In the case of one-dimensional deterministic overdamped ratchets chaotic behavior has been observed by avoiding  the Poincar\'e-Bendixon theorem \cite{Zarlenga2009}. For example,  adding stochasticity to an overdamped ratchet with quenched disorder is one way to obtain chaos and anomalous diffusion in the system \cite{Popescu2000}. Long range spatially correlated quenched disorder also produces anomalous diffusion in overdamped ratchets and both the amount of
quenched disorder and the degree of correlation  can enhance the anomalous diffusive transport \cite{Gao2003}.

Synchronization is a phenomenon of considerable scientific and technological interest \cite{Synchronization2012}.   In the case of ratchets, synchronized motion of particles with an external
sinusoidal driving force has been studied  for both a perfect
and a disordered ratchet potential \cite{Zarlenga2005,Zarlenga2007}. In the disordered ratchet potential, anomalous diffusion  was associated with a new trapping mechanism \cite{Zarlenga2005,Zarlenga2007}.
Coupling overdamped ratchets increases the order of the dynamical equations and, as a consequence chaos may be obtained \cite{Cillia2001,Fendrik2006,Goko2005,Wang2007,Vincent2010}. 
Another strategy to avoid the conditions of Poincar\'e-Bendixon theorem is to use a discontinuous periodic driving
force so that the vector field is no longer $C^1$. Chaos and multiple synchronization
becomes possible because trajectories for non-$C^1$ fields
may be discontinuous. This approach was considered in \cite{Zarlenga2009} where a deterministic overdamped ratchet driven by a periodic square driving force was  shown to display chaotic
behavior. The  strong nonlinearity of the driving force
produces a bifurcation pattern with synchronized as well as chaotic regions.
The necessary and sufficient conditions that the ratchet
potential under a periodic square-wave driving force must satisfy in order to have a vanishing current were obtained by
Salgado-Garc\'ia et al. \cite{Salgado2006}. Recently,  the first experimental
realization of a deterministic optical rocking ratchet under a periodic square-wave driving force was obtained by Arzola et al. \cite{Arzola2011}. A periodic and
asymmetric light pattern was made to interact with dielectric microparticles in water, giving rise to a
ratchet potential. The motion of the microparticles with respect to the pattern with an unbiased time-periodic square-wave
function tilts the potential in alternating opposite directions. A thorough analysis of the dynamics of the system and a comparison between theoretical and experimental results are presented in \cite{Arzola2013}.

Our main interest in this paper is to gain a deeper insight into the chaotic behavior of deterministic overdamped ratchet
systems
by considering positive and negative delta functions as the unbiased driving force. The reason for using alternate positive and negative pulses as driving force is twofold, on the one hand the integrability of the dynamical equations  makes it possible to obtain analytical maps of the particle dynamics, and on the other hand, due to the strong nonlinearity of the delta function the system develops a 
rich dynamical behavior.

We show in this work that alternate positive and negative delta functions as the unbiased driving force on a ratchet potential produces  both
synchronized and chaotic regions. Being an autonomous one-dimensional (1-D) system, it is possible to obtain a 1-D map where the transition from regular to chaotic motion can be studied. We show that a tangent bifurcation diagram is associated with this transition by analytically obtaining a 1-D map and  studying the corresponding power spectrum.
In order to investigate the dynamics of the system with other driving forces, we consider  a group of $n$ positive delta function pulses followed by no force up to the end of the first half of the period $T$ and then the same number of negative pulses with a time interval with no force up to the end of the full period $T$.  The number of pulses $n$ is then increased until it fills the entire time interval $T$.  We compare the dynamics of this system with the case of a continuous square wave as the driving force used in \cite{Zarlenga2009}. In both cases the synchronization regions are equivalent, showing that the continuous driving force may be considered as a succession of delta function pulses.

The outline of the paper is as follows. In Sec. \ref{sec:rat} we present the
kicked-ratchet model and discuss the synchronization regions and the  bifurcation diagram.  The analytical map is derived in
Sec. \ref{sec:map}. In Sec. \ref{sec:periodic} we compare the case of a  continuous square wave with the delta function built pulse. Finally, conclusions are presented in Sec.\ref{sec:conclusions}.

\section{The kicked ratchet: transport and synchronization}
\label{sec:rat}
The model under study is an overdamped ratchet, where non-interacting particles  move through a ratchet potential, under a viscous friction with coefficient $\gamma$. The particles are driven by a periodic force $f_T(t)$. The dynamical equation for each particle is as follows:
\begin{equation}
\label{eq:ecuacion}
\gamma \dot{x}=R_{\lambda}(x)+f_T(t).
\end{equation}
The  ratchet force $R_{\lambda}(x)$ is periodic in $x$ with spatial period $\lambda$, and it has  zero spatial mean value, $\langle R_{\lambda}(x)\rangle _x=0$. The conservative ratchet force is related to the ratchet potential $U(x)$ by the relation,
\begin{equation}
\label{eq:fuerzaratchet}
R_{\lambda}(x)=-dU/dx,
\end{equation}
where $U(x)$ is analytically defined by:
\begin{equation}
\label{eq:potencialratchet}
U(x)=-A[\sin (2\pi x/\lambda)+\frac{\mu }{2}\sin(4\pi x/\lambda)].
\end{equation}
To make contact with our previous work  \cite{Larrondo2003} we use  $A=1$,  $\lambda=2\pi$,   $\gamma=0.1109$ , and $\mu=0.5$.

The driving force $f_{T}(t)$, is an alternating periodic sequence of positive and negative delta functions with weight $\pm J$ and period $T$. It may be expressed as follows:
\begin{equation}
\label{eq:deltatrain}
f_T(t)=\sum_{i=1}^{\infty}(-1)^iJ\delta(t-iT/2).
\end{equation}
This driving force has zero temporal mean value $\langle  f_T(t) \rangle _t=0$. Then the complete model under study is:
\begin{equation}
\label{eq:ecuacionNor}
\dot{x}=\frac{cos(x) + 0.5 cos(2x)}{\gamma}+\frac{J}{\gamma}\sum_{i=1}^{\infty}(-1)^i\delta(t-iT/2),
\end{equation}
with $J $ and $T $  as control parameters. Relevant scales are the spatial period of the ratchet $\lambda$ for coordinate $x$, the period of the external force $T$ for the time $t$ and $v_{\omega}=\lambda / T$ for velocities.
The advantage of using delta function pulses  is to minimally disturb the system letting it to  evolve free, between each pulse. Consequently it is possible to understand the dynamics by only analyzing the fixed points and the characteristic times of the autonomous system.

We integrated Eq. \ref{eq:ecuacionNor} using a fourth order, variable step,  Runge Kutta algorithm  with $\Delta T=0.005$ and $\Delta J=0.005$. The studied region of the parameter space is $0.8< T< 2.0$,  and $0< J< 20$. For  $J=0$ and $T=0.8$ the system starts at $x=x_{max}\simeq -1.19$. This value corresponds to a maximum of the ratchet potential $U(x)$.  For the other values of $T$ and $J=0$ the initial condition is the final value for the previous $T$, reduced to the first well.  For the next value of $J=0.005$ and the starting value of $T=0.8$ the initial condition is the final value for $J=0$ and $T=0.8$, reduced to the first well. For the other values of $T$ and $J=0.005$ the initial condition is the final value for the previous $T$,  and so on. This is the \textit{method II} used in ref. \cite{Larrondo2002}. For each value of $J$ and $T$, the initial data for  a transitory time  of $t_{tran}=100T$ is discarded and then the trajectory starts to be sampled before each positive delta function pulse and it is stored from $t=100T$ to  $t=160T$.

Two aspects of each trajectory are considered: the rotation number of the oscillations and the current through the ratchet.  Let $q$ be the number of periods of the driving force required for the state variable $x$  and its derivative $v=\dot{x}$ to repeat within an error $err_x \le 0.01\lambda$ and $err_v=0.01 v_{\omega}$ respectively. This $q$ is in fact the denominator of the \textsl{rotation number} \cite{Larrondo2003}. The synchronization is verified up to $q=60$.
The current is measured in  units of the mean velocity $\langle v\rangle =({x_{160T}-x_{100T}})/{60T}$.

\begin{figure}
\begin{tabular}{cc}
\centerline{\includegraphics[width=1\textwidth]{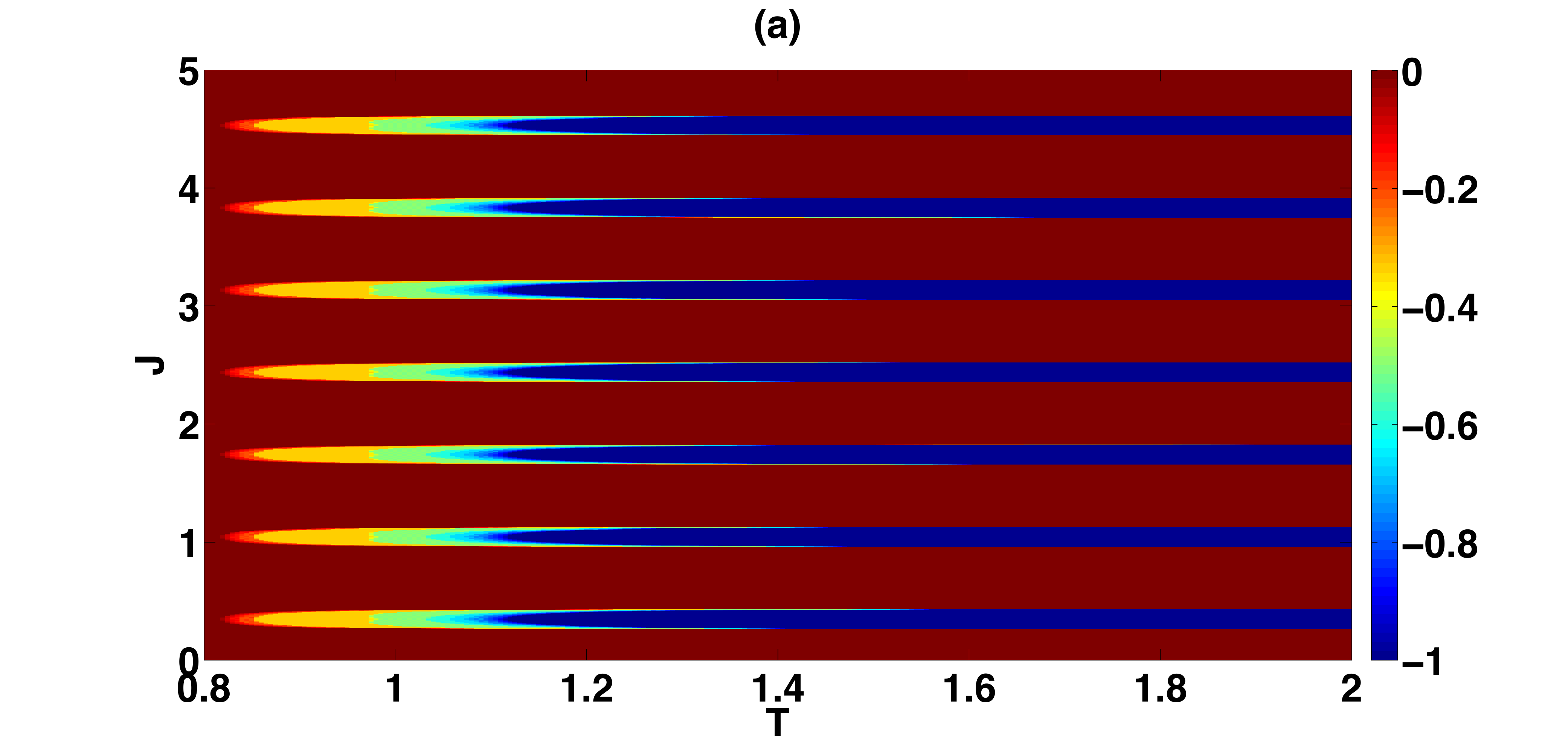}}\\
\centerline{\includegraphics[width=1\textwidth]{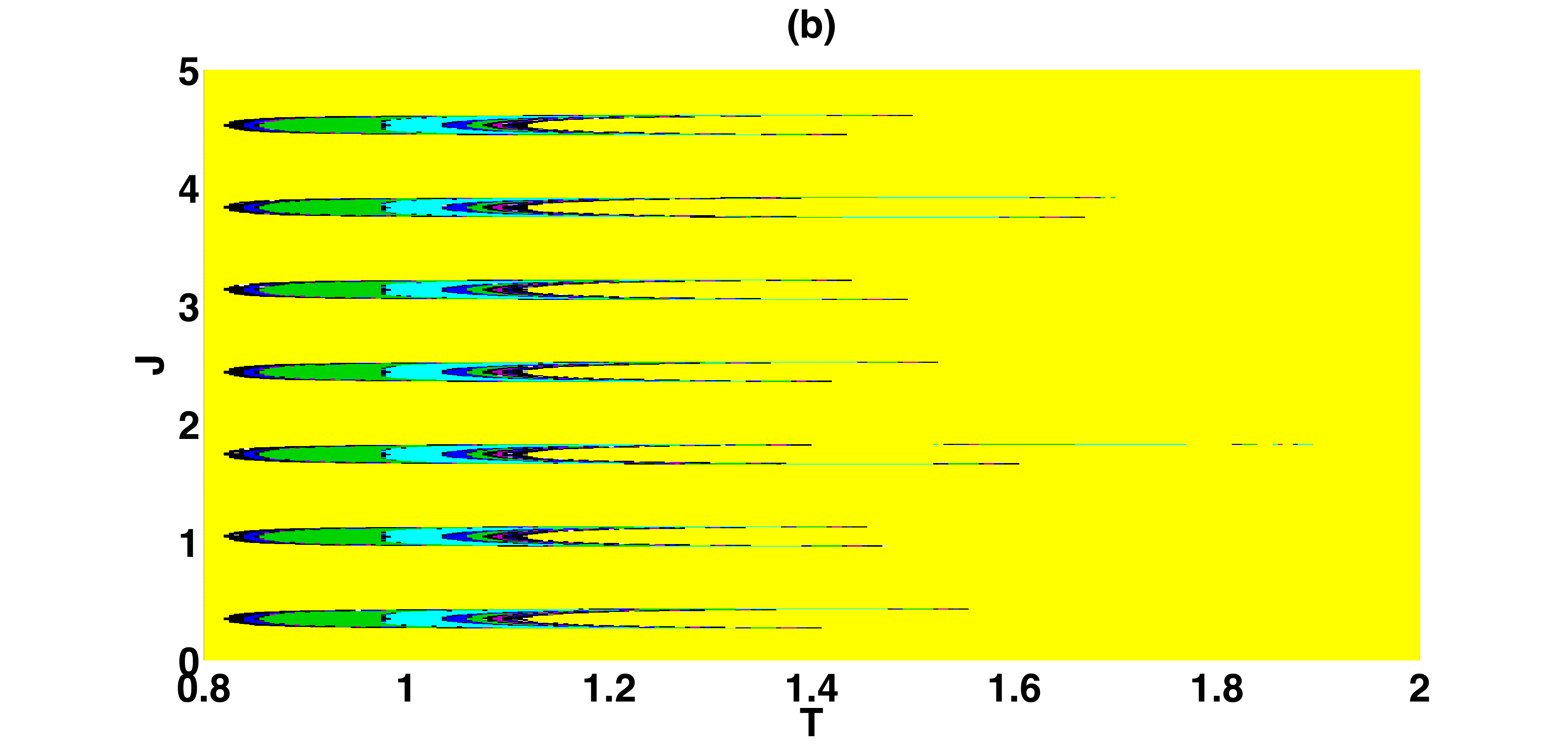}}
\end{tabular}\caption{(a) The mean velocity $\langle v\rangle $, and (b) values of $q$ are plotted in the $J-T$ parameter space,  with $\gamma=0.1109$ and $\mu=0.5$. The indicator bar on the right in (a) shows the corresponding values of the mean velocity. The color scale is as follows: \textit{cyan} for $q=1$, \textit{blue} for $q=2$, \textit{green} for $q=3$,          \textit{orange} for $q=4$, \textit{magenta} for $q=5$, \textit{red} for $6\le q < 32$, and \textit{yellow} for $q > 32$. }
\label{fig:globales}
\end{figure}

The results for the normalized mean velocity, $\langle \tilde{v} \rangle=\langle v \rangle /v_\omega$, and the $q$ values, in the $J-T$ parameter space, are shown in Figures \ref{fig:globales}(a) and \ref{fig:globales}(b), respectively, for $0.8< T< 2.0$, and $0< J< 5$.  The color scheme is described in the figure caption.

Comparing Figs. \ref{fig:globales}(a) and  \ref{fig:globales}(b) clearly shows that the particle oscillates with $\langle v\rangle  =0$  and $q=1$ for most values of $T$ and $J$. But there are \textsl{stripe-like} regions with interesting transport properties, where $\langle v\rangle \neq 0$ and different integer values of $q$ are possible. The mean velocity is always negative meaning that particles only move backwards.
The maximum value of the modulus of the mean velocity is $|\langle v\rangle /v_{\omega}|=1$, which corresponds to synchronized regions in the $J-T$ parameter space with  $q=1$.

\begin{figure}
\begin{tabular}{cc}
\centerline{\includegraphics[width=1\textwidth]{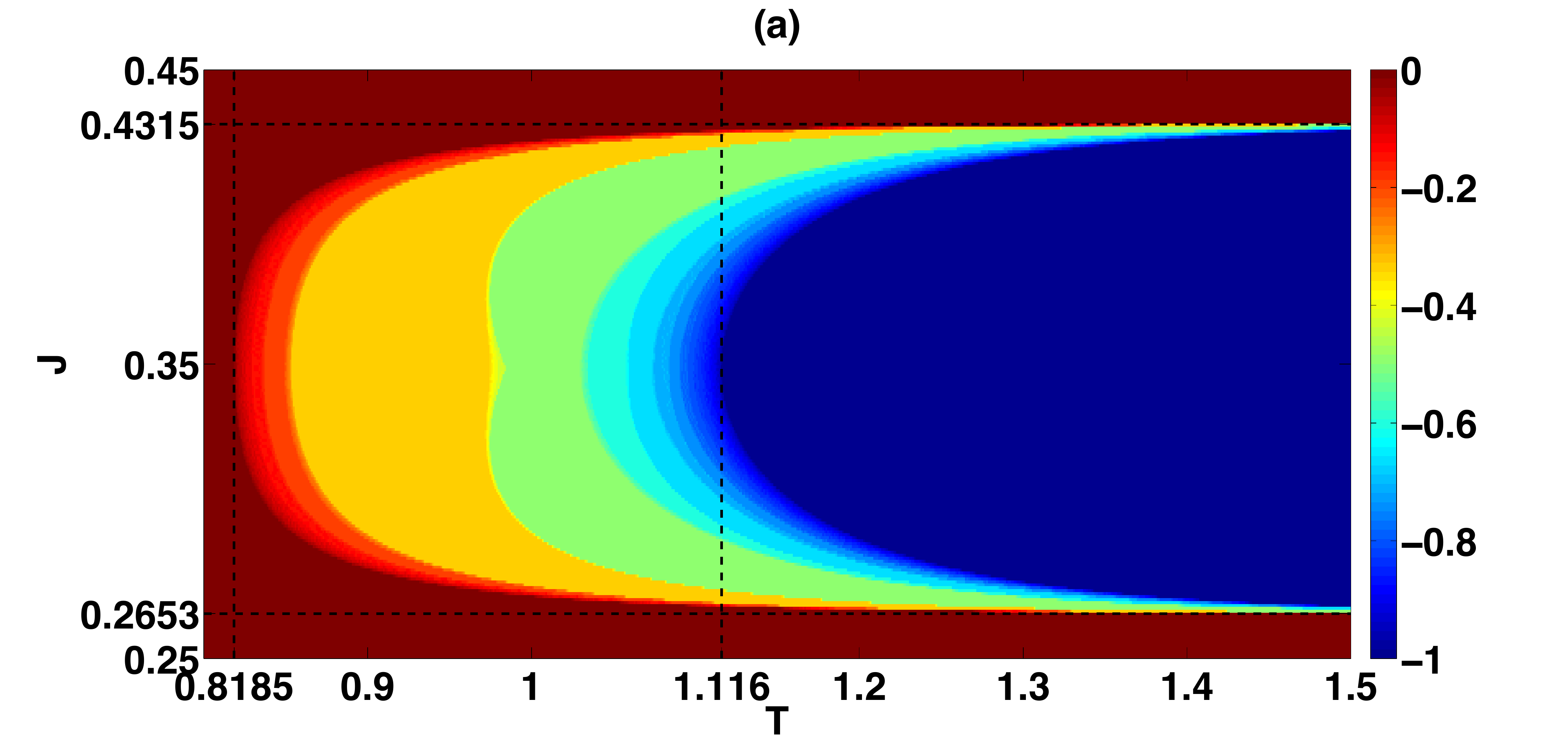}}\\
\centerline{\includegraphics[width=1\textwidth]{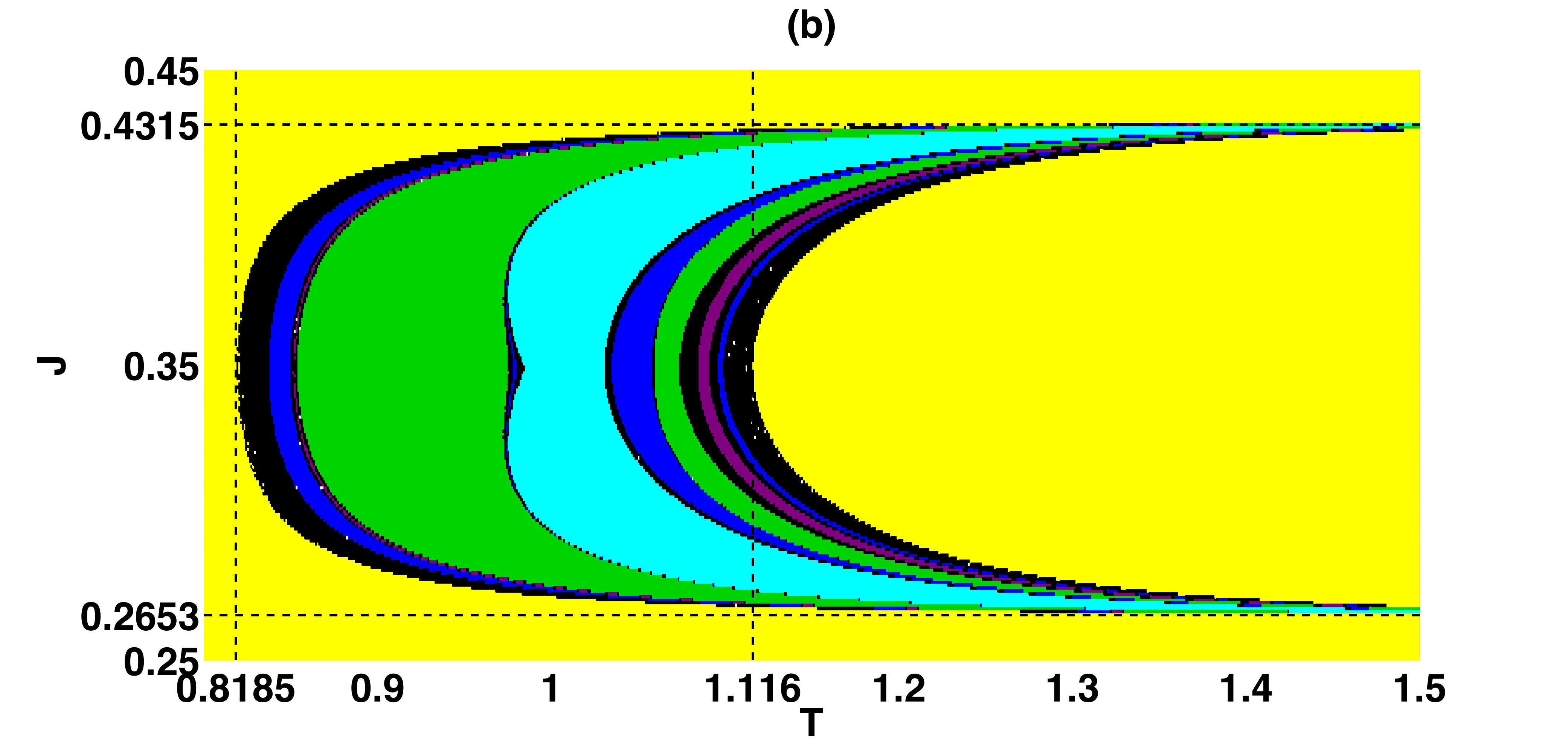}}
\end{tabular}
\caption{(a) Enlarged views of the mean velocity $\langle  v\rangle $, and (b) values of $q$ in the first stripe in Fig. \ref{fig:globales} are plotted in the $J-T$ parameter space,  with $\gamma=0.1109$ and $\mu=0.5$. The indicator bar on the right in (a) shows the corresponding values of the mean velocity. The color scale is: \textit{yellow (lightest gray)} for $q=1$, \textit{cyan (second lightest gray)} for $q=2$, \textit{green (third lightest gray)} for $q=3$, \textit{purple (fourth lightest gray)} for $q=4$, \textit{blue (darkest gray)} for $q=5$, \textit{black} for $6\le q \le 32$, and \textit{white} for $q > 32$.  }
\label{fig:zoom}
\end{figure}

In Figs. \ref{fig:zoom}(a) and \ref{fig:zoom}(b) we show  enlarged views of one of the stripes in Fig. \ref{fig:globales}. The horizontal dotted lines in the figures mark the values $J=0.2653$, and $J=0.4315$. The vertical dotted lines mark the values $T=0.8185$  and $T=1.116$.  The significance of these regions are discussed below.

\begin{figure}
\begin{tabular}{cc}
\centerline{\includegraphics[width=1\textwidth]{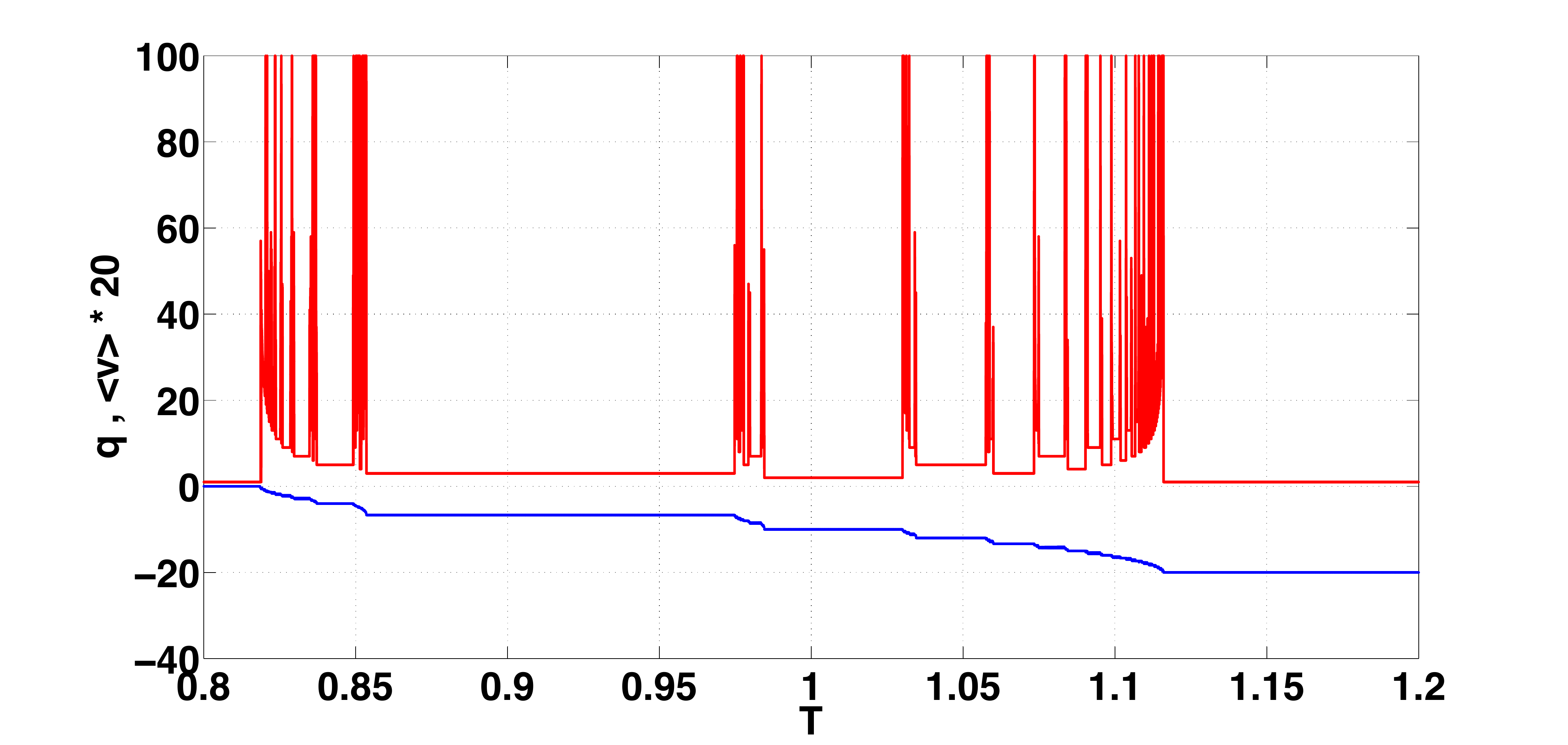}}
\end{tabular}
\caption{Bifurcation diagram showing the dependence of the mean velocity (red) and the values of $q$ (blue) as a function of $T$, with $J=0.3485$, $\gamma=0.1109$ and $\mu=0.5$. }
\label{fig:bifu}
\end{figure}

The  bifurcation diagrams for $\langle \tilde {v}\rangle $ and $q$ as functions of  $T$, with $J=0.3485$, are shown in Fig. \ref{fig:bifu}. In these figures synchronization is found with a higher precision than in Figs. \ref{fig:globales} and \ref{fig:zoom}, using errors $\tilde{err_x}=0.001 $ and $\tilde{err_ v}=0.001$, respectively.

We can give a simple explanation for the origin of the stripes in Fig. \ref{fig:globales}. Fig. \ref{fig:fijosYsaltos} shows the ratchet potential as a function of the dimensionless units $x/\lambda$.  As usual, in 1-D systems fixed points are alternatively stable and unstable. The location of stable fixed points is $\tilde {x}^*_s=x^*_s/\lambda= a\simeq0.19036+n$ and unstable fixed points are located at $\tilde {x}^*_u=x^*_u/\lambda= -a\simeq-0.19036+n$. Domains of attraction of the stable fixed points are limited by non-stable ones. If the starting position of a particle, $\tilde{x}_0$, is in the domain of attraction of the stable fixed point $\tilde{x}^s_i$, without external forcing the particle  will move through the ratchet to this stable fixed point.

A delta function with weight $\pm J$ forces the particle to jump from $\tilde{x}_0$ to $\tilde{x}_0\pm J/(\gamma\lambda)$. This new position may be in a different domain of attraction than the initial position. If this is the case, the particle evolves inside a different well. Suppose the particle starts at the stable fixed point $\tilde {x}^*_s\simeq 0.19036$. The necessary condition for a positive current is $J/(\gamma\lambda) > 1-2a\simeq 0.6192$. For $\lambda=2\pi$ and $\gamma$=0.1109, this implies that $J\gtrsim 0.4315$. Similarly, a negative current requires that $J/(\gamma\lambda) > 2a\simeq 0.3807$. For $\lambda=2\pi$ and $\gamma$=0.1109, this implies that $J\gtrsim0.2653$. These values are  shown as dotted lines in Fig. \ref{fig:zoom}.  They correspond to  the limiting values of each stripe zone. The negative current is favored, because it requires a smaller value of $J$ to reach the domain of attraction of a stable fixed point to the left, rather than to the right. Note that for  $0.2653< J< 0.4315$ the negative  $\langle \tilde{v}\rangle $ has a maximum absolute value of $1$, because the positive delta function pulse does not change the well of the particle but the negative one does. This analysis is exact if $T$ is high enough to allow the particle to reach a stable fixed point between the delta function pulses. The value $T=1.1034$ shown in Figs. \ref{fig:zoom} is the characteristic time over which the particle does reach the stable fixed point between delta function pulses within $\tilde{err_x}$.

For  $0.4315\lesssim J \lesssim1.2653$ a positive delta function force moves the particle one well forward and the negative one moves the particle one well backward. Consequently, $\langle v\rangle =0$ again.

If $J$ is increased to $J/(\gamma\lambda)>  1+2a$, meaning $J\gtrsim1.2653$ a new stripe region appears. In this region,  for the same $T$,  the particle has the same values of $\langle v\rangle $ as in the previous stripe, but $\langle v\rangle /v_{\omega}=-1$, for example.  This implies that the particle goes forward one well and backwards two wells in each period $T$.

\section{Analytical map}
\label{sec:map}

\begin{figure}
\begin{tabular}{cc}
\includegraphics[width=0.5\textwidth]{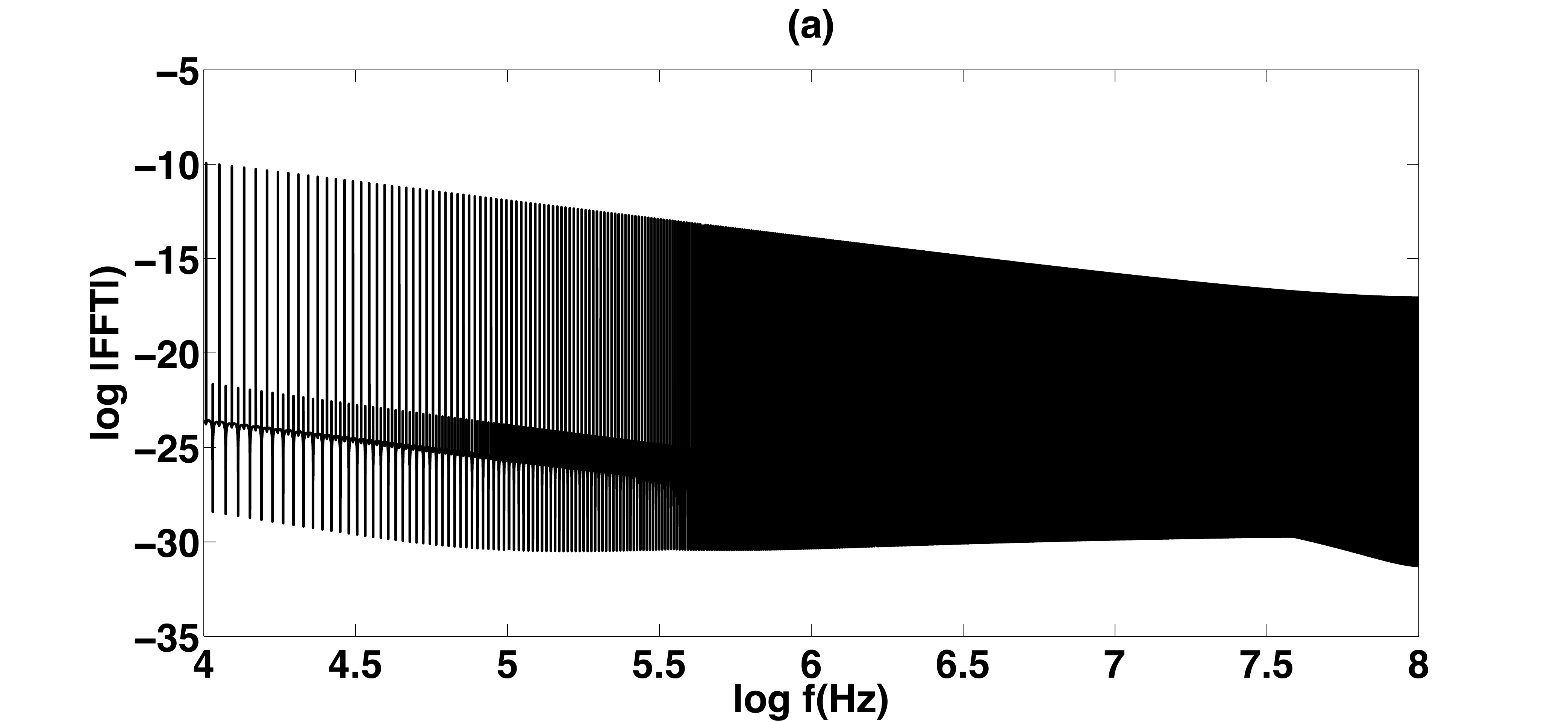}
\includegraphics[width=0.5\textwidth]{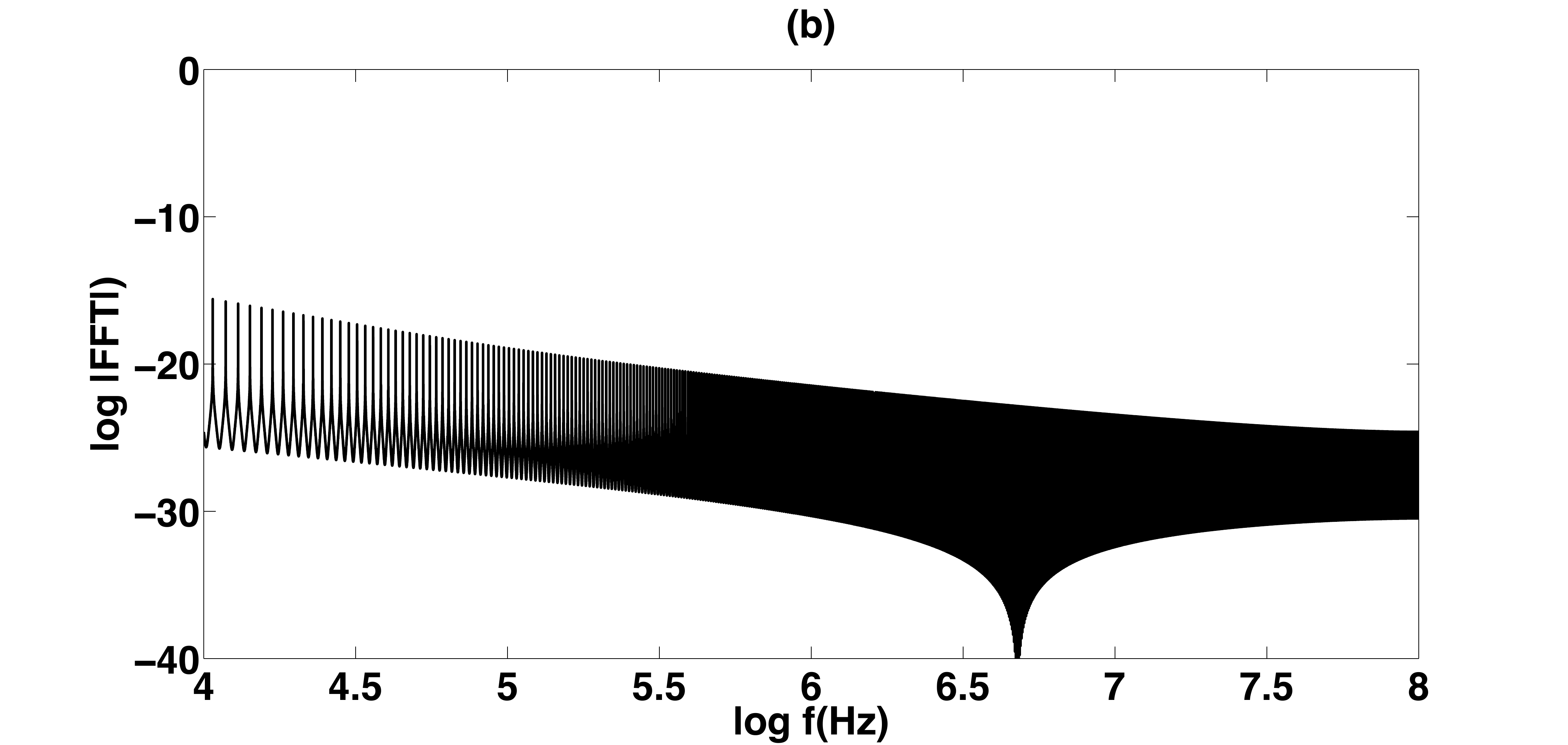}
\end{tabular}
\caption{Power spectrum for the the time series $\{x_n\}$ generated by an overdamped ratchet with $\gamma=0.1109$ and $\mu=0.5$, for the same bifurcation points in parameters space of Fig. \ref{fig:bifusmapas} (note that the slope in log-log scale is $-2$: (a)  $T=0.81850$ and $J=0.3485$; (b) $T=1.11600$ and $J=0.3485$. }.
\label{fig:psk}
\end{figure}

In order to analyze the dynamical behavior of the particles near the border between zones with different $q$ and $\langle v\rangle $,
  the time series  for $T=0.85115$ and $J=0.3485$ is chosen, because for these values the system displays a small negative mean velocity and a high value of $q$ (see Figs. \ref{fig:zoom}). Fig. \ref{fig:psk} shows the power spectrum of this time series. The spectrum shows the characteristic $1/f$ power-law behavior, corresponding to an instability produced by the tangent bifurcation. 

The autonomous system is $1$-D and consequently it is possible to construct a $1$-D map to analyze the transition from regular motion with a rational value for $v/v_{\omega}$, to an irregular chaotic region with an irrational value of $v/v_{\omega}$.  Below, we list the steps in the derivation of the map.

\begin{enumerate}
\item Suppose at $t=0$ the particle is at the position $x_n$. Applying a positive delta function pulse forces the particle  to jump to position $x_1$. As a result, we can write,
\begin{equation}
\label{eq:Ts2}
\int_{x_n}^{x_1}{dx}=\frac{1}{\gamma}\int_{0^-}^{0^+}{[cos(x)+\mu cos(2x)+J]dt}=\frac{J}{\gamma}.
\end{equation}
Where,
\begin{equation}
\label{eq:saltomas}
x_1=x_n+\frac{J}{\gamma}
\end{equation}
\item After the force is applied, the system evolves following the autonomous differential equation:
\begin{equation}\label{eq:autonomo}
\gamma\dot{x}=cos(x)+\mu cos(2x).
\end{equation}
Eq. \ref{eq:autonomo} may be integrated as follows:
\begin{eqnarray}\label{eq:tmas}
\nonumber t^{+}(x) &=& \int_{x_n}^{x}{\frac{\gamma}{cos(x) + \mu cos(2x)}dx}\\
& =& \left. \frac{2^{3/2}}{3^{3/4}} \left\{tanh^{-1}\left[\frac{1+\sqrt{3}}{12^{1/4}}tan(x/2)\right]-tan^{-1}\left[\frac{-1+\sqrt{3}}{12^{1/4}}tan(x/2)\right]
\right\} \right |_{x_1}^{x}
\end{eqnarray}

At the end of the first half-period $t=T/2$  the particle reaches position $x_{1/2}$ which is the solution of:

\begin{eqnarray}\label{eq:0Ts2}
\nonumber \frac{T}{2}=\frac{2^{3/2}}{3^{3/4}} \left\{tanh^{-1}\left[\frac{1+\sqrt{3}}{12^{1/4}}tan(\frac{x_{1/2}}{2})\right]-tan^{-1}\left[\frac{-1+\sqrt{3}}{12^{1/4}}tan(\frac{x_{1/2}}{2})\right]  \right.\\
\left. -tanh^{-1}\left[\frac{1+\sqrt{3}}{12^{1/4}}tan(\frac{x_{1}}{2})\right]+tan^{-1}\left[\frac{-1+\sqrt{3}}{12^{1/4}}tan(\frac{x_{1}}{2})\right]\right\}
\end{eqnarray}

\item Now, a negative delta function force is applied and the particle jumps to position $x_2$, where,

\begin{equation}\label{eq:saltomenos}
x_2=x_{1/2}-\frac{J}{\gamma}
\end{equation}
\item From then on, the dynamical equation is again Eq. \ref{eq:autonomo}, and during the interval $\Delta t=T/2$ the particle finally reaches position $x_{n+1}$ which is the solution of the following equation:

\begin{eqnarray}
\label{eq:ts2T}
\nonumber \frac{T}{2}=\frac{2^{3/2}}{3^{3/4}} \left\{tanh^{-1}\left[\frac{1+\sqrt{3}}{12^{1/4}}tan(\frac{x_{n+1}}{2})\right]-tan^{-1}\left[\frac{-1+\sqrt{3}}{12^{1/4}}tan(\frac{x_{n+1}}{2})\right]  \right.\\
\left. -tanh^{-1}\left[\frac{1+\sqrt{3}}{12^{1/4}}tan(\frac{x_{2}}{2})\right]+tan^{-1}\left[\frac{-1+\sqrt{3}}{12^{1/4}}tan(\frac{x_{2}}{2})\right]\right\}
\end{eqnarray}
\end{enumerate}
Then $x_{n+1}=f(x_n)$ is the solution of the system of equations \ref{eq:0Ts2} to \ref{eq:ts2T}.

\begin{figure}
\begin{tabular}{cc}
\includegraphics[width=0.5\textwidth]{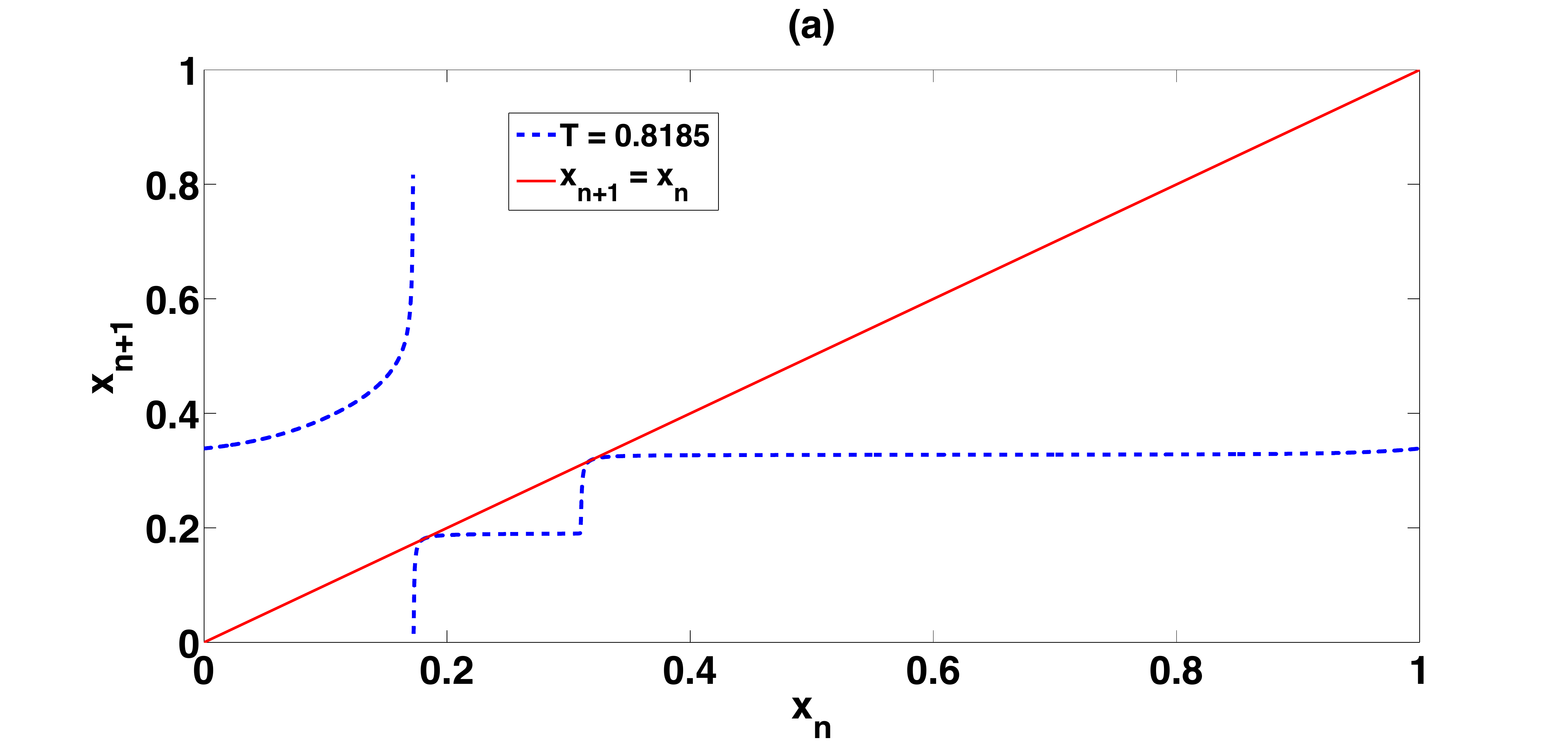}
\includegraphics[width=0.5\textwidth]{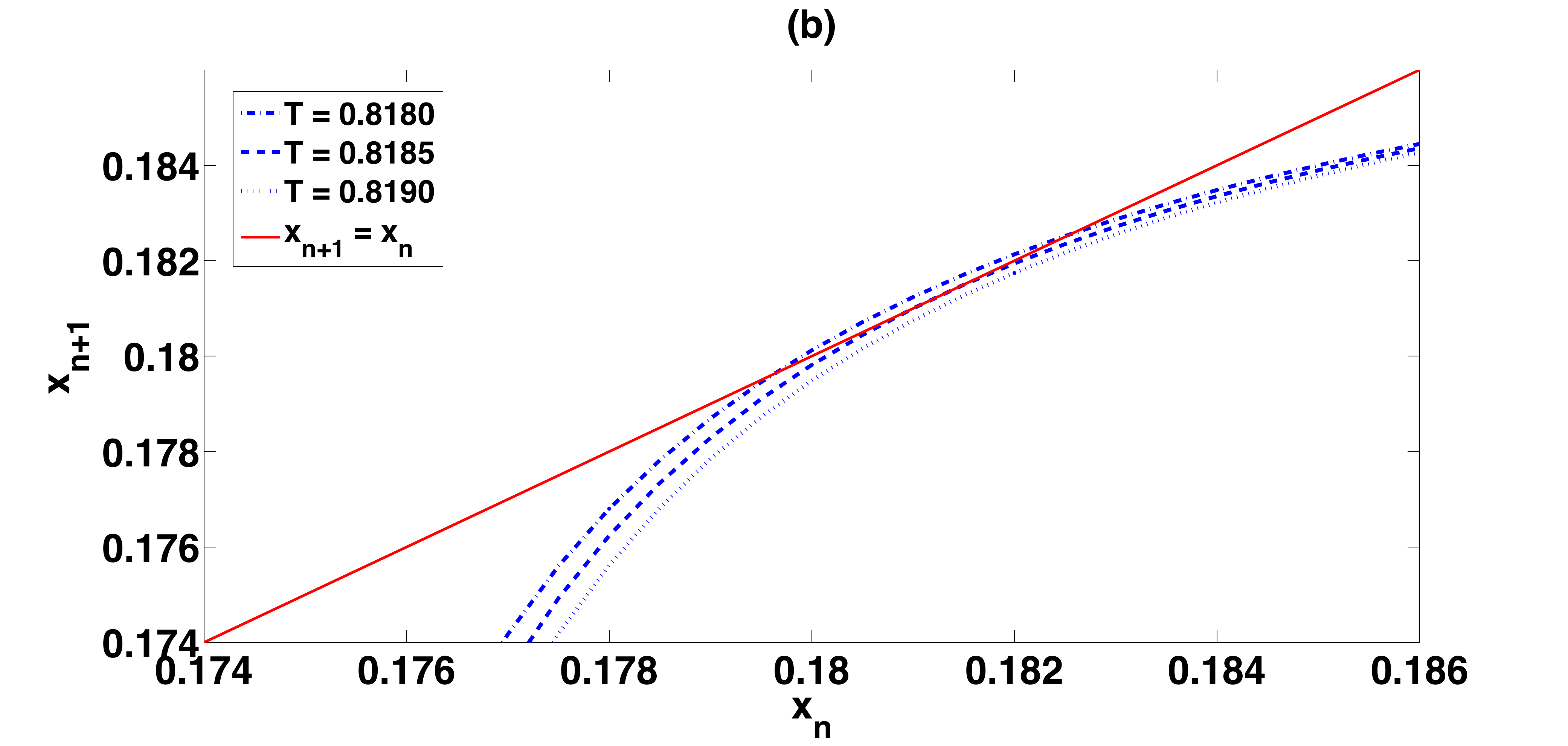}\\
\includegraphics[width=0.5\textwidth]{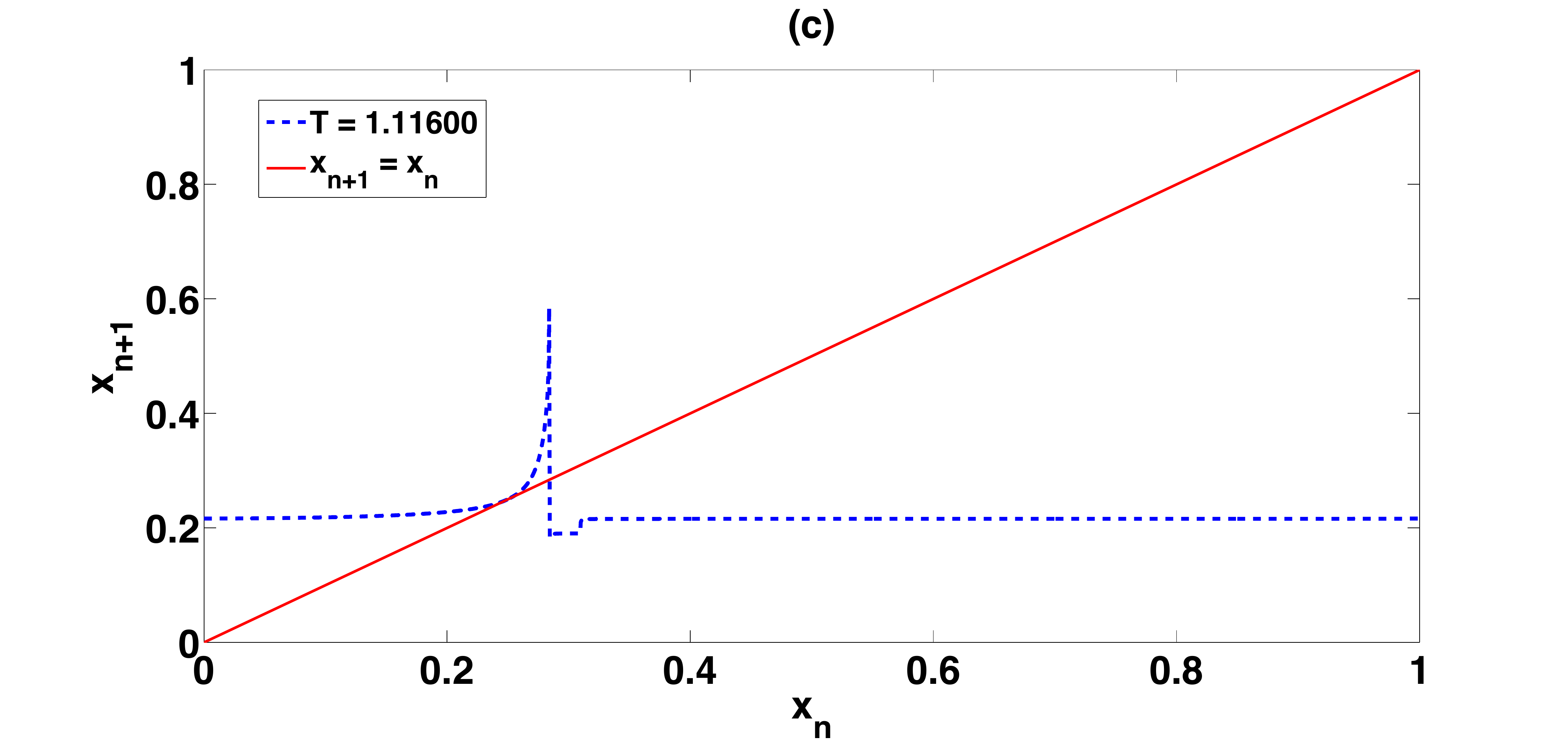}
\includegraphics[width=0.5\textwidth]{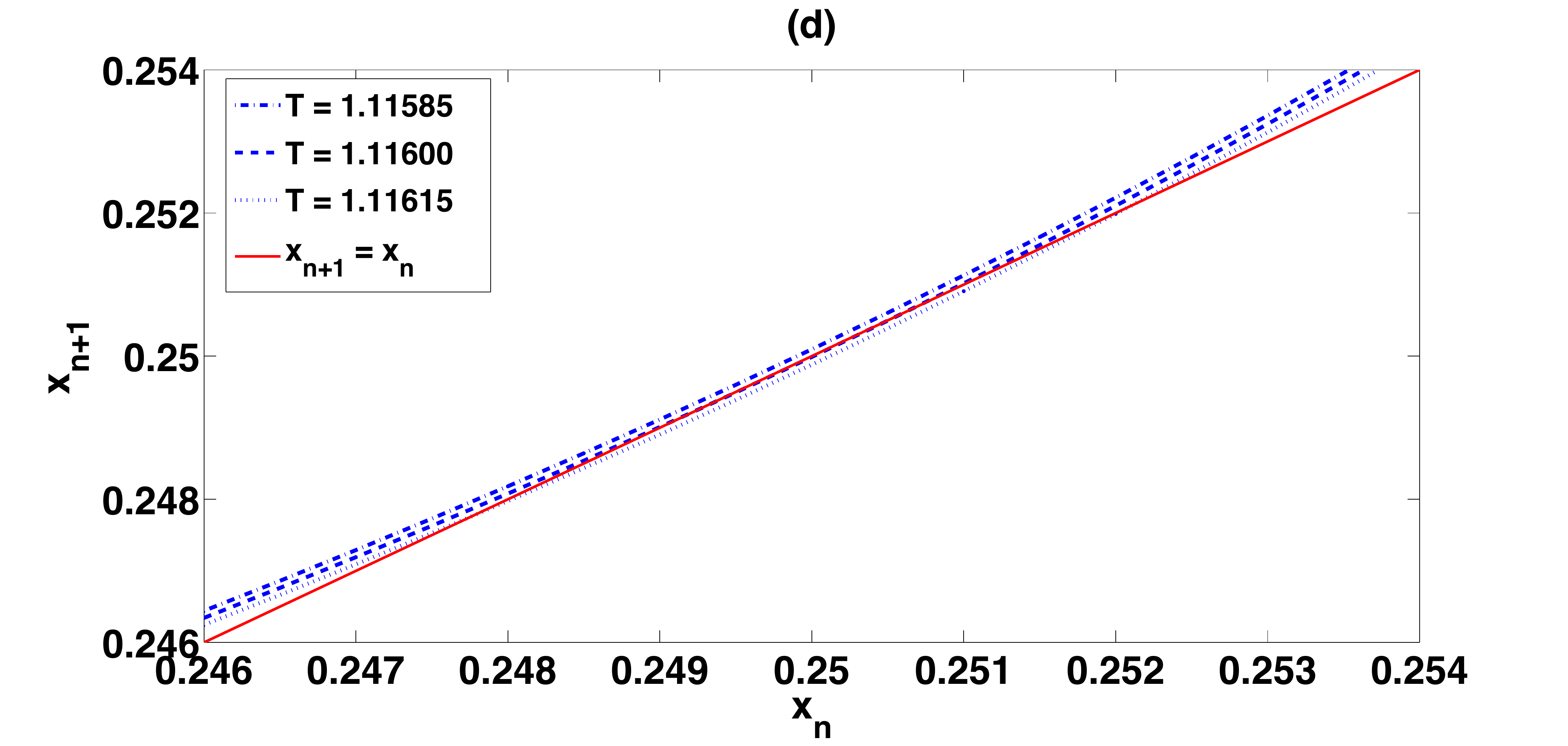}
\end{tabular}
\caption{One dimensional map of the system showing the tangent bifurcation in two points of the parameter's space. The system is the  overdamped ratchet with $\gamma=0.1109$ and $\mu=0.5$. (a)$T\simeq 0.81850$ and $J=0.3485$; (b) Zoom of Fig. (a) near the bifurcation point; $T\simeq 1.11600$ and $J=0.3485$; (c) Zoom of Fig. (c) near the bifurcation point.}
\label{fig:bifusmapas}
\end{figure}

Figs. \ref{fig:bifusmapas}(a) and  \ref{fig:bifusmapas}(c) show the analytical map for $T=0.81850$ and $T=1.11600$, respectively, for $J=0.3485$. These parameter values are inside the chaotic regions (see Fig. \ref{fig:bifu}). 
Figs. \ref{fig:bifusmapas}(b) and \ref{fig:bifusmapas}(d) show, respectively, zooms of maps in (a) and (c) showing that the tangent bifurcation is the origin of the instability.  The fact that the bifurcation is the origin of the instability is also confirmed by the $1/f^2$ behavior of the power spectra shown in Figs. \ref{fig:psk}(a) and \ref{fig:psk}(b). 

\section{Other periodic forces}
\label{sec:periodic}

\begin{figure}
\begin{tabular}{cc}
\includegraphics[width=0.4\textwidth]{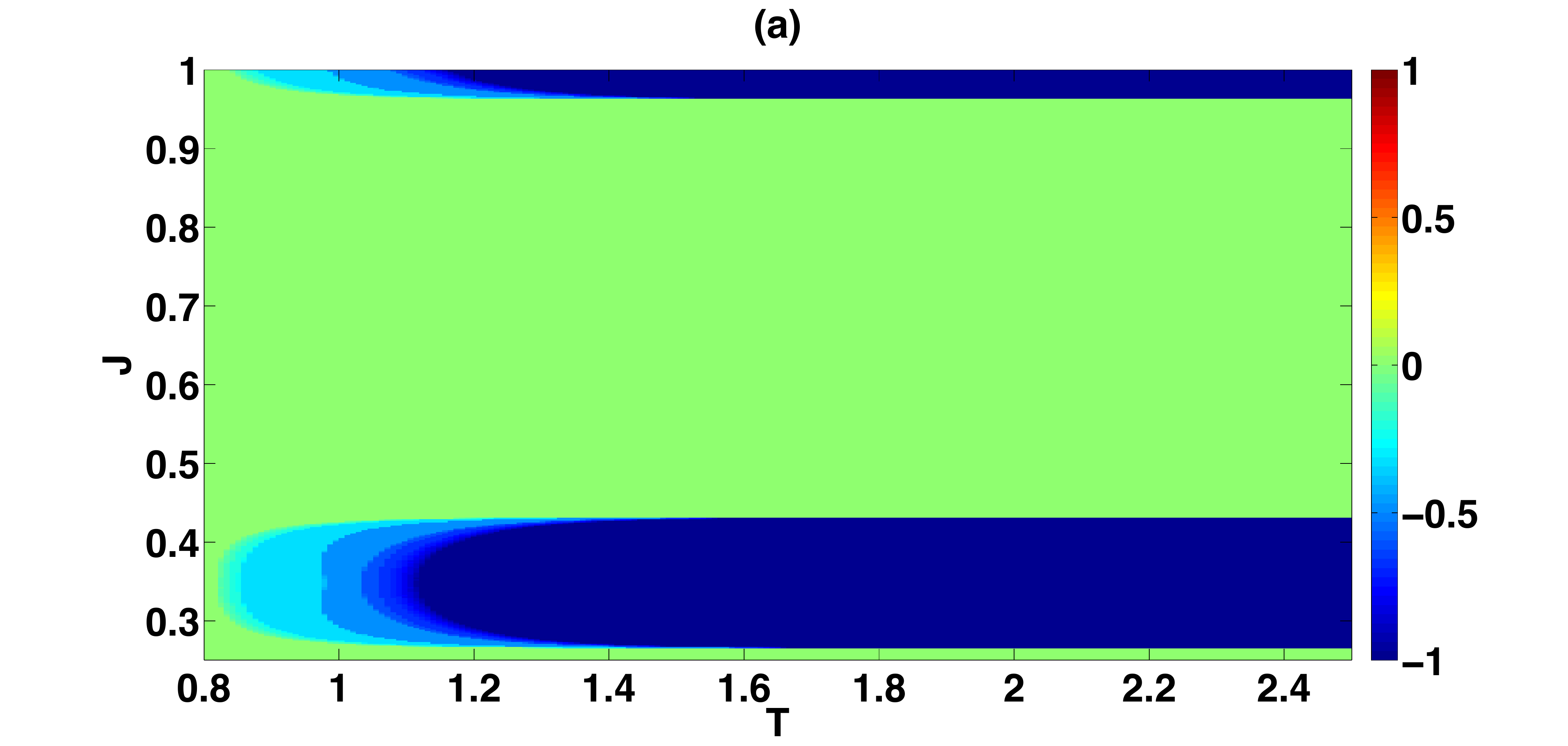}
\includegraphics[width=0.4\textwidth]{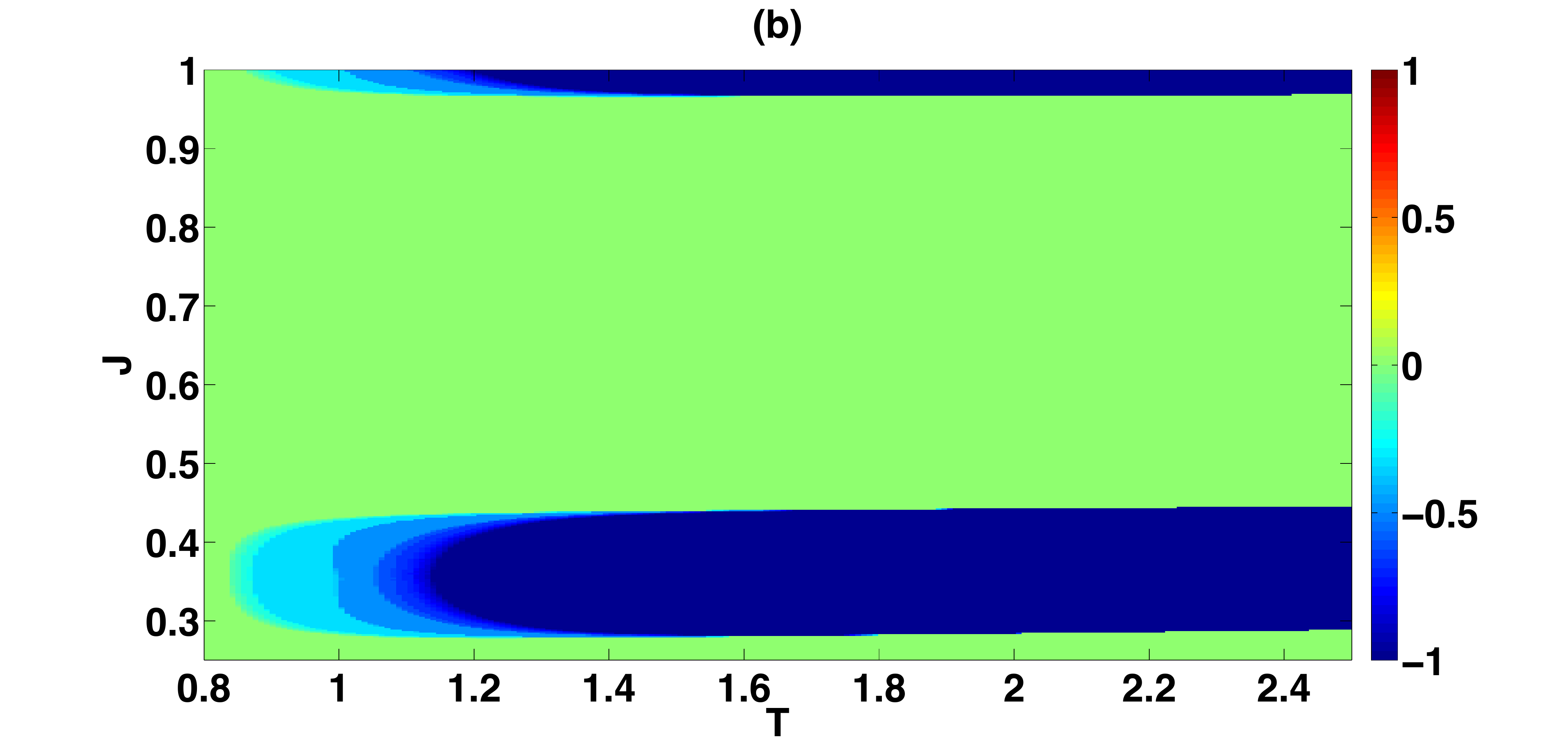}\\
\includegraphics[width=0.4\textwidth]{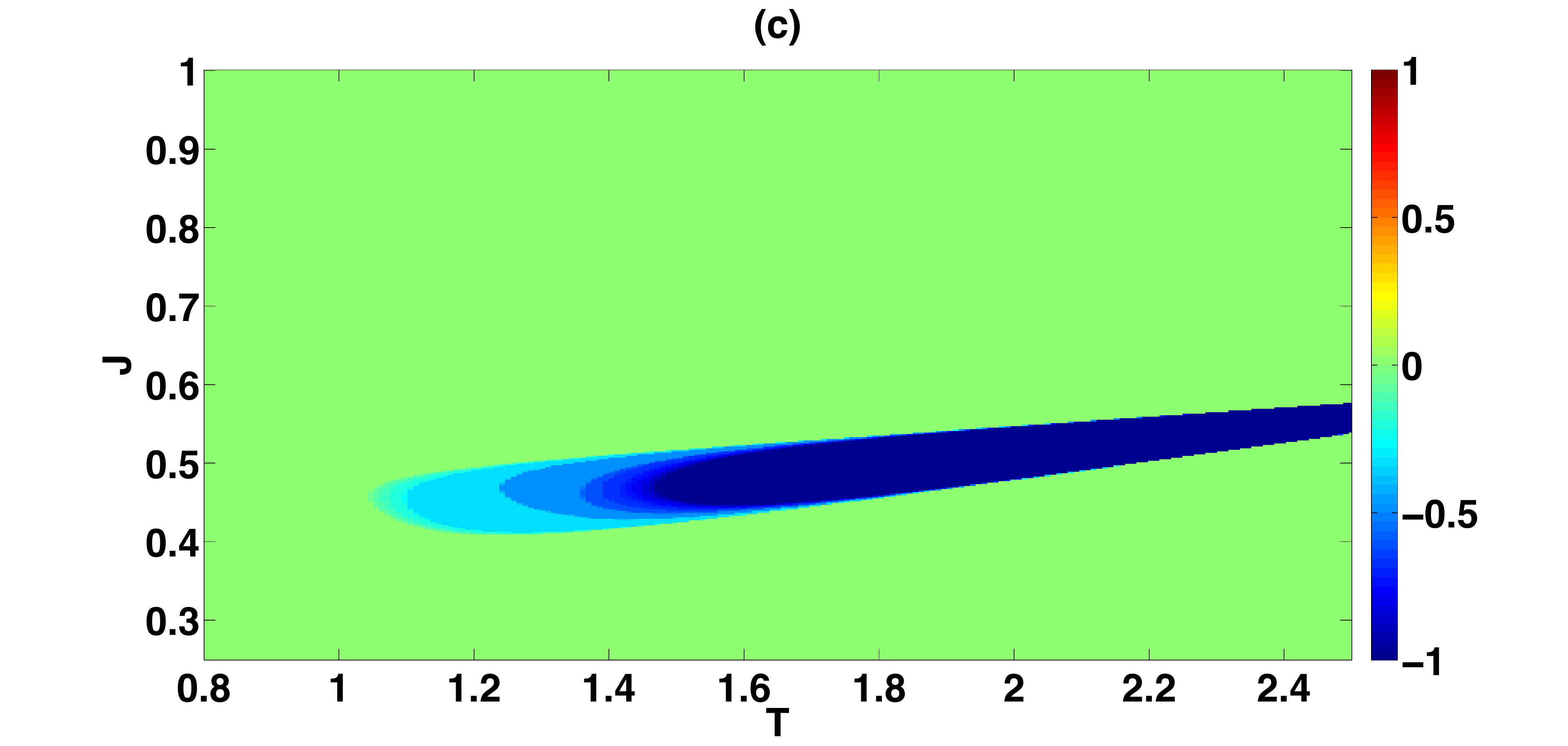}
\includegraphics[width=0.4\textwidth]{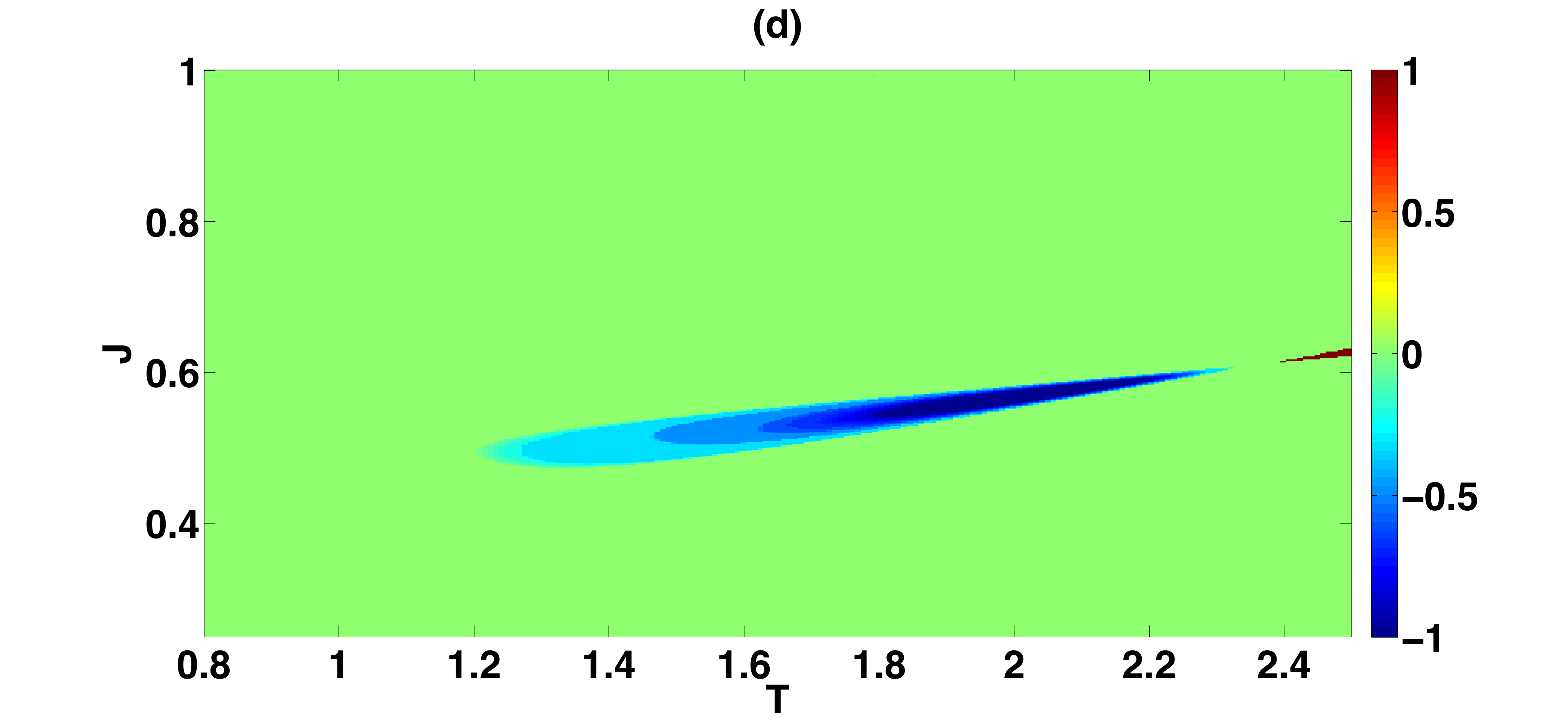}\\
\includegraphics[width=0.4\textwidth]{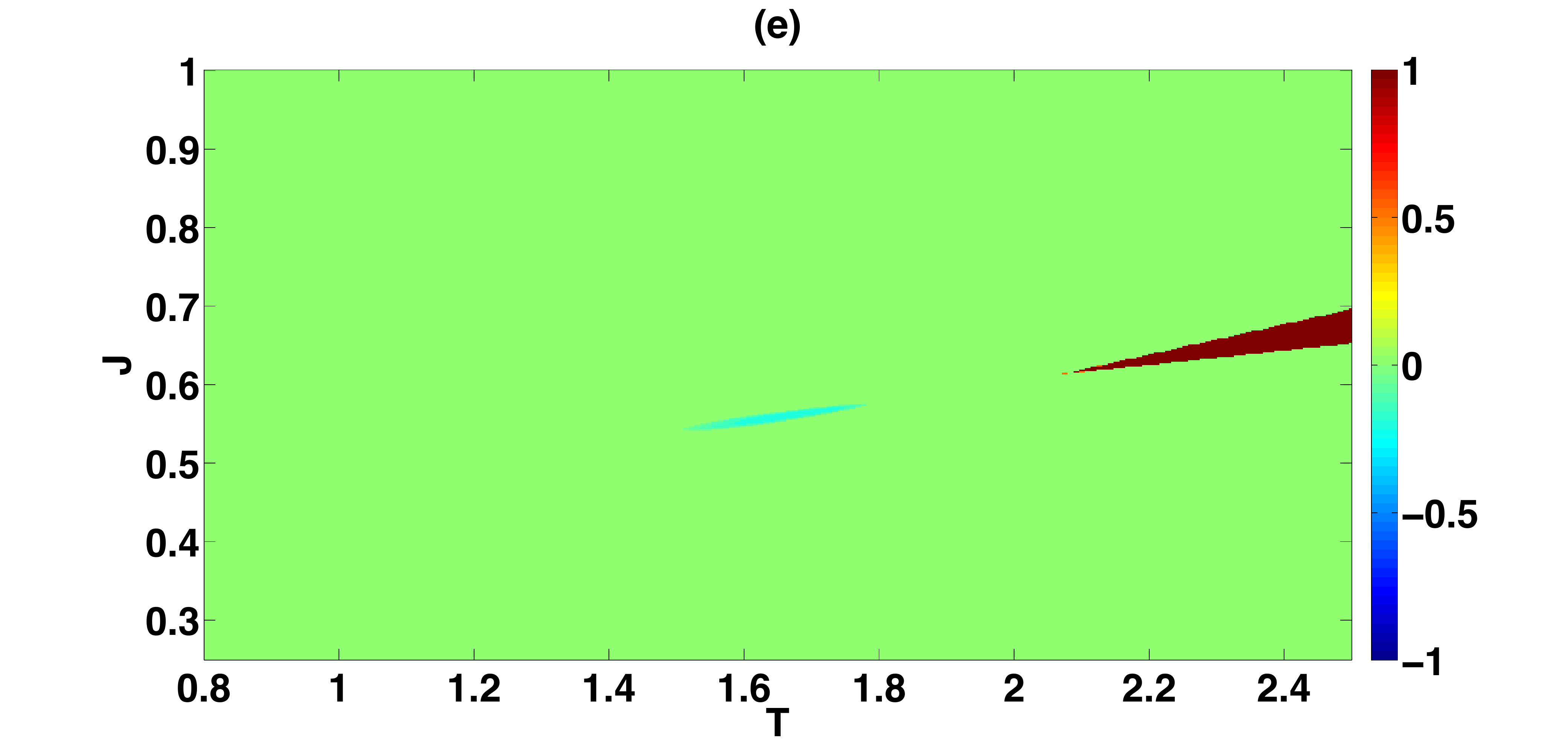}
\includegraphics[width=0.4\textwidth]{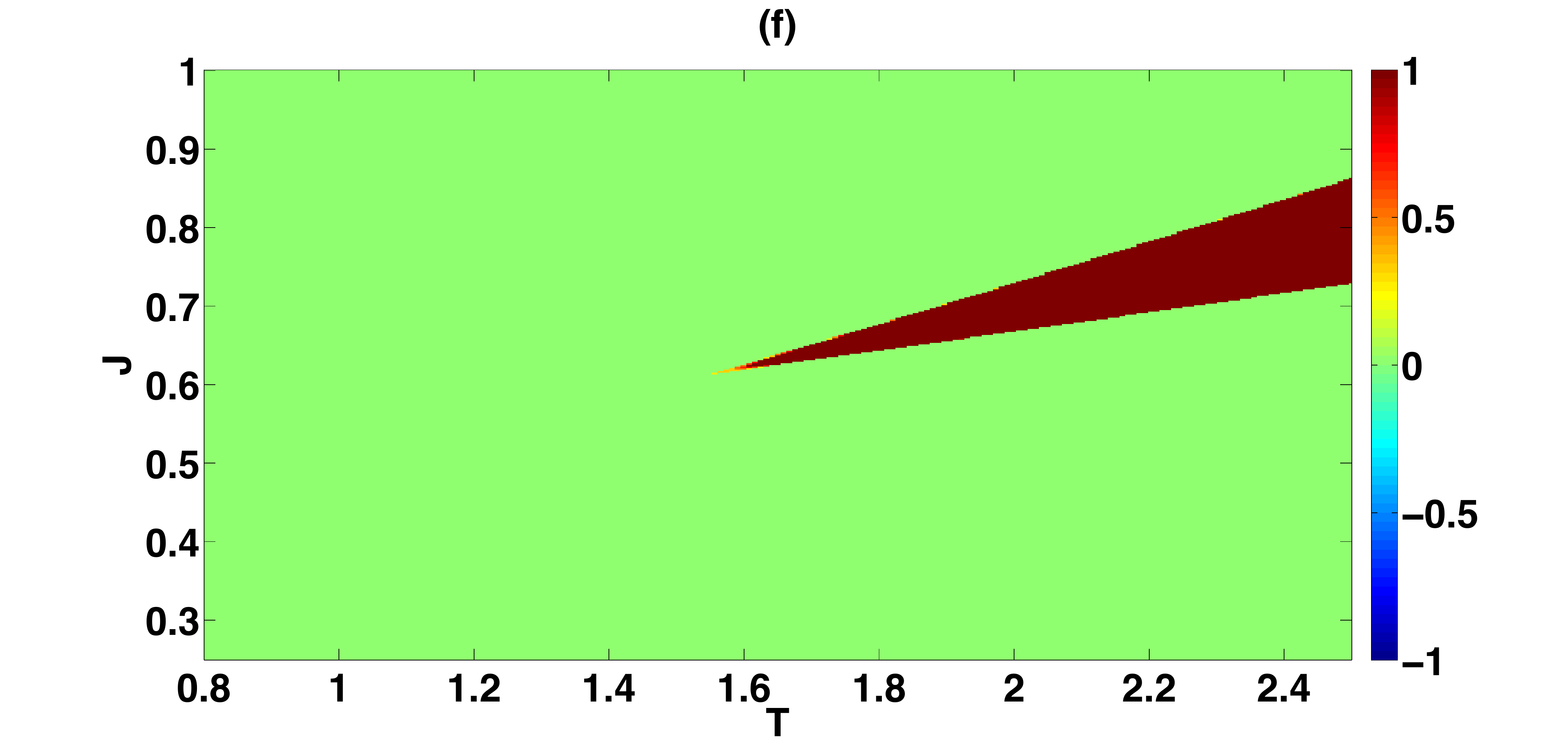}\\
\includegraphics[width=0.4\textwidth]{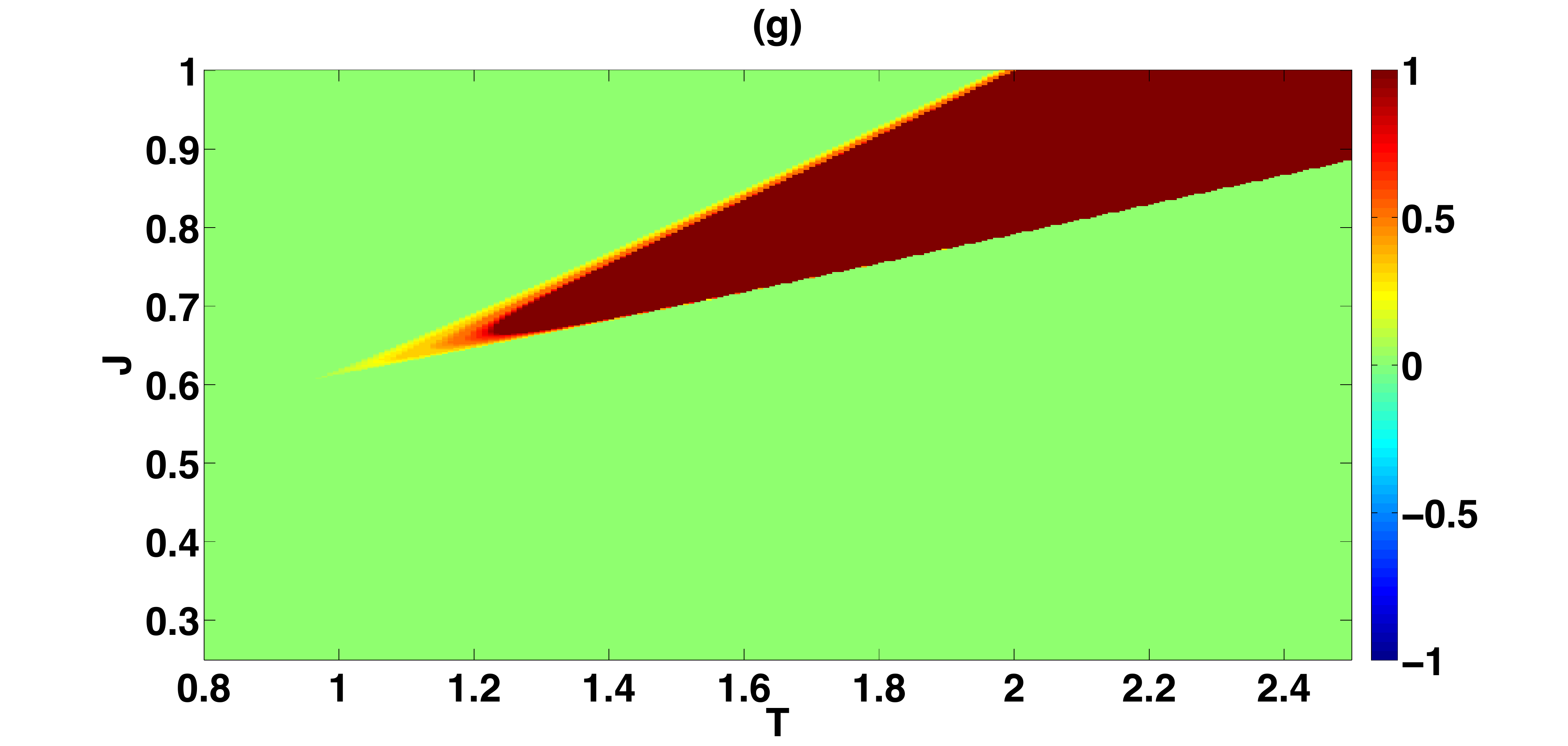}
\includegraphics[width=0.4\textwidth]{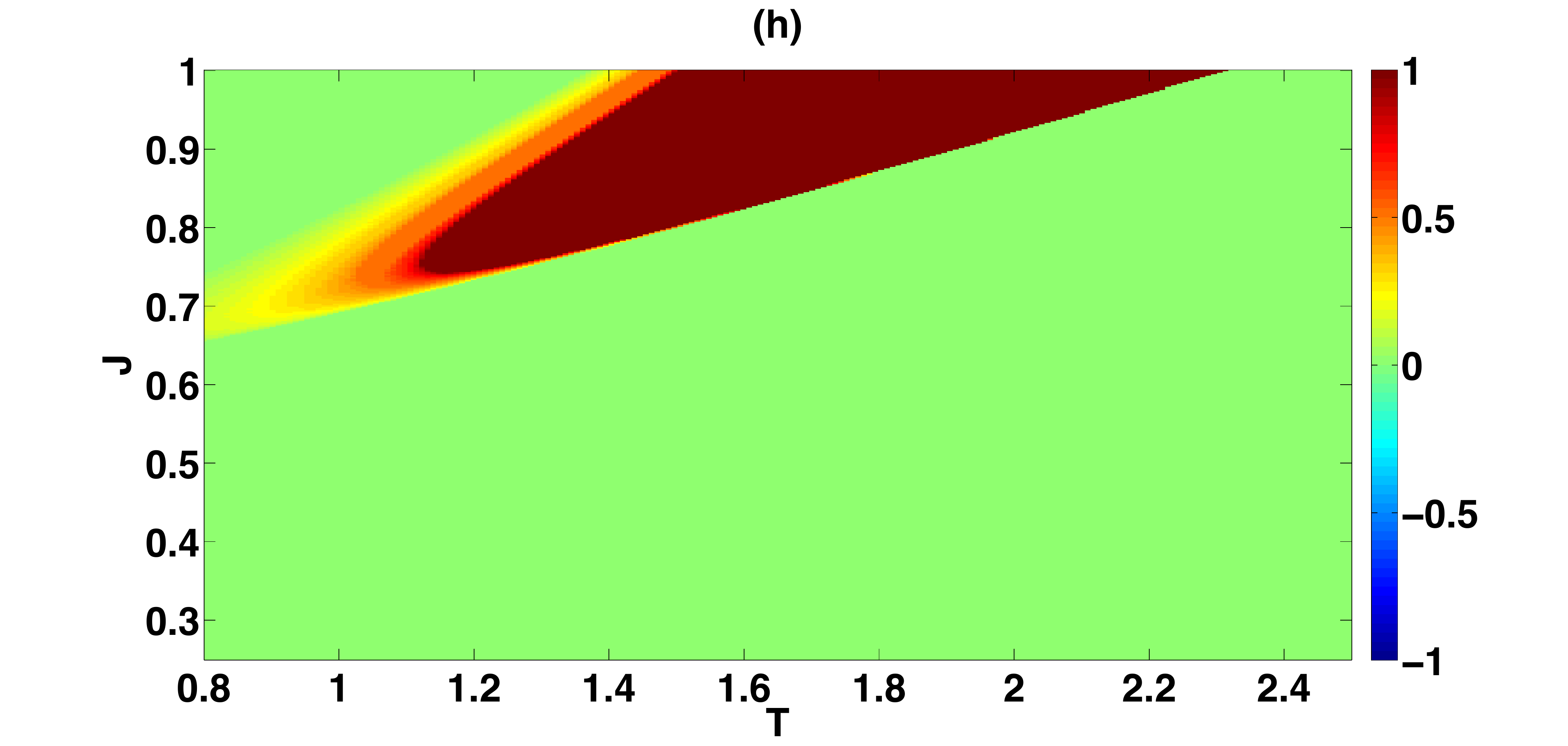}\\
\end{tabular}
\caption{Bifurcation regions of $\langle v\rangle $ in the parameter's space $[J,T]$ for positive and negative delta trains of length $K$. (a) $K=1$; (b) $K=10$; (c) $K=100$ ;(d)$K=130$ ; (e) $K=150$; (f) $K=200$; (g) $K=300$; (h) $K=400$;}.
\label{fig:avvel}
\end{figure}

It is interesting to extend the above analysis to other more general types of periodic forces.  Specifically, consider the case of $K$ consecutive positive deltas  applied at times $t=0+iT/1000+nT$, with $i=0$ to $K-1$, and the same number $K$ of negative delta function pulses at times $t=T/2+iT/1000+nT$. To produce the same total impulse for all $K$, each pulse has a weight $J_i=J/K$.  Figs. \ref{fig:avvel}(a) to \ref{fig:avvel}(h) show the bifurcation behavior of the mean velocity in the $J-T$ parameter space  for different values of $K$. There are two \textsl{idle times} after the positive deltas and the negative deltas, given by $t_{iddle}=T/2-(K-1)~T/1000$. 

\begin{figure}
\begin{tabular}{cc}
\centerline{\includegraphics[width=1\textwidth]{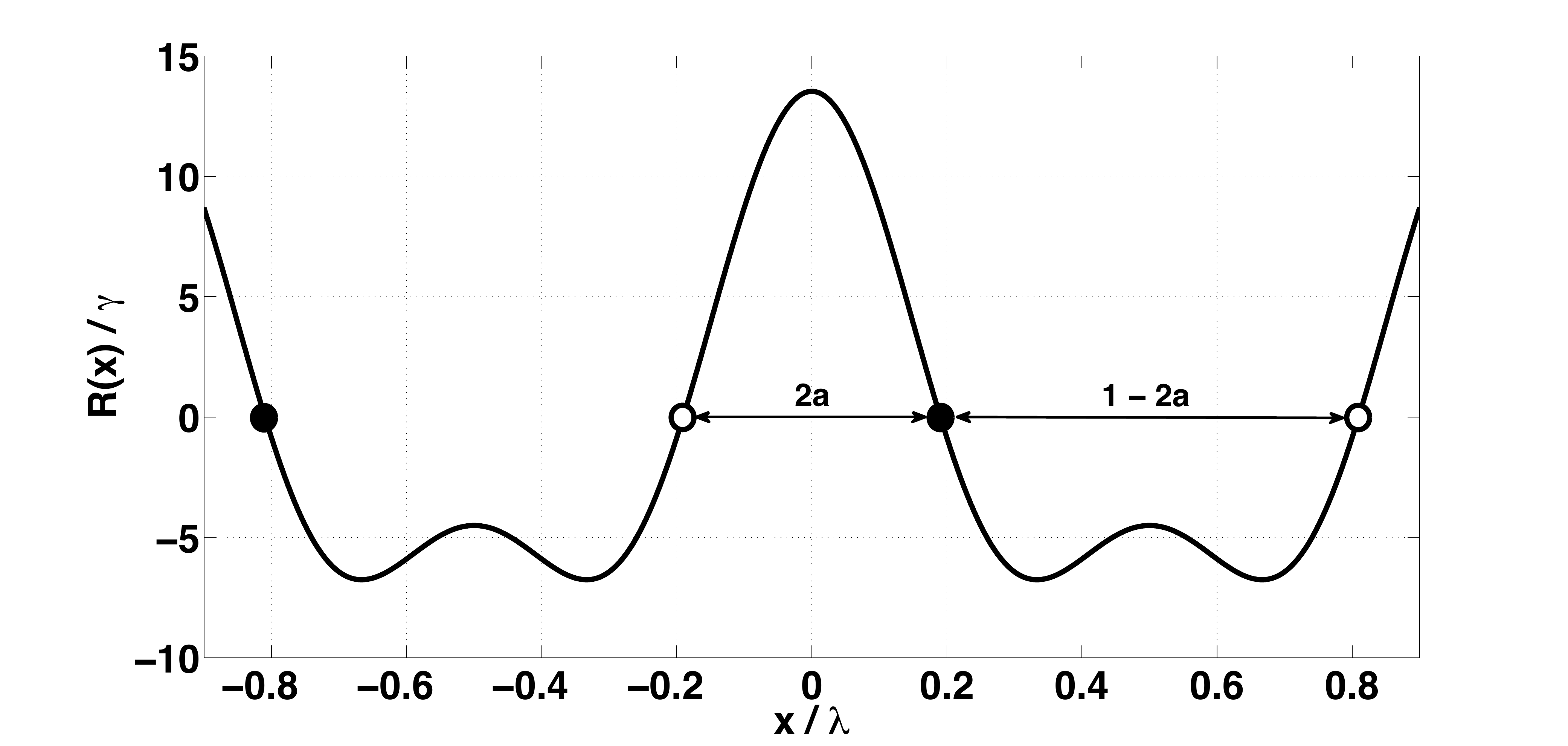}}
\end{tabular}
\caption{The ratchet potential with fixed points for $\gamma=0.1109$ and $\mu=0.5$. Stable fixed points are located at $\tilde{x}^*_s=a+n, n\in \mathbb{Z}$ with $a\simeq 0.1904$. Unstable fixed points are located at $\tilde{x}^*_u=-a + n,n\in \mathbb{Z}$.  }.
\label{fig:fijosYsaltos}
\end{figure}

The sequence in Fig. \ref{fig:avvel} shows the disappearance of the synchronization region with negative transport properties and the appearance of another synchronization region with positive transport, as $K$ increases. The diagram remains almost identical to that obtained for $K=1$ if the number $K$  of delta function pulses is small. In that case  the analysis using the one dimensional map explained in Section \ref{sec:map} remains valid. But if the number of delta function pulses increases, and the duty cycle $\alpha=(K-1)/500$ increases, the synchronization regions with negative current disappear. Progressively a new region with positive current emerges and the one dimensional analytical map is no longer valid. Let us now consider an intermediate $K=130$ value in order to gain a deeper insight in this issue (see Fig. \ref{fig:fijosYsaltos})

\begin{figure}
\begin{tabular}{cc}
\includegraphics[width=0.8\textwidth]{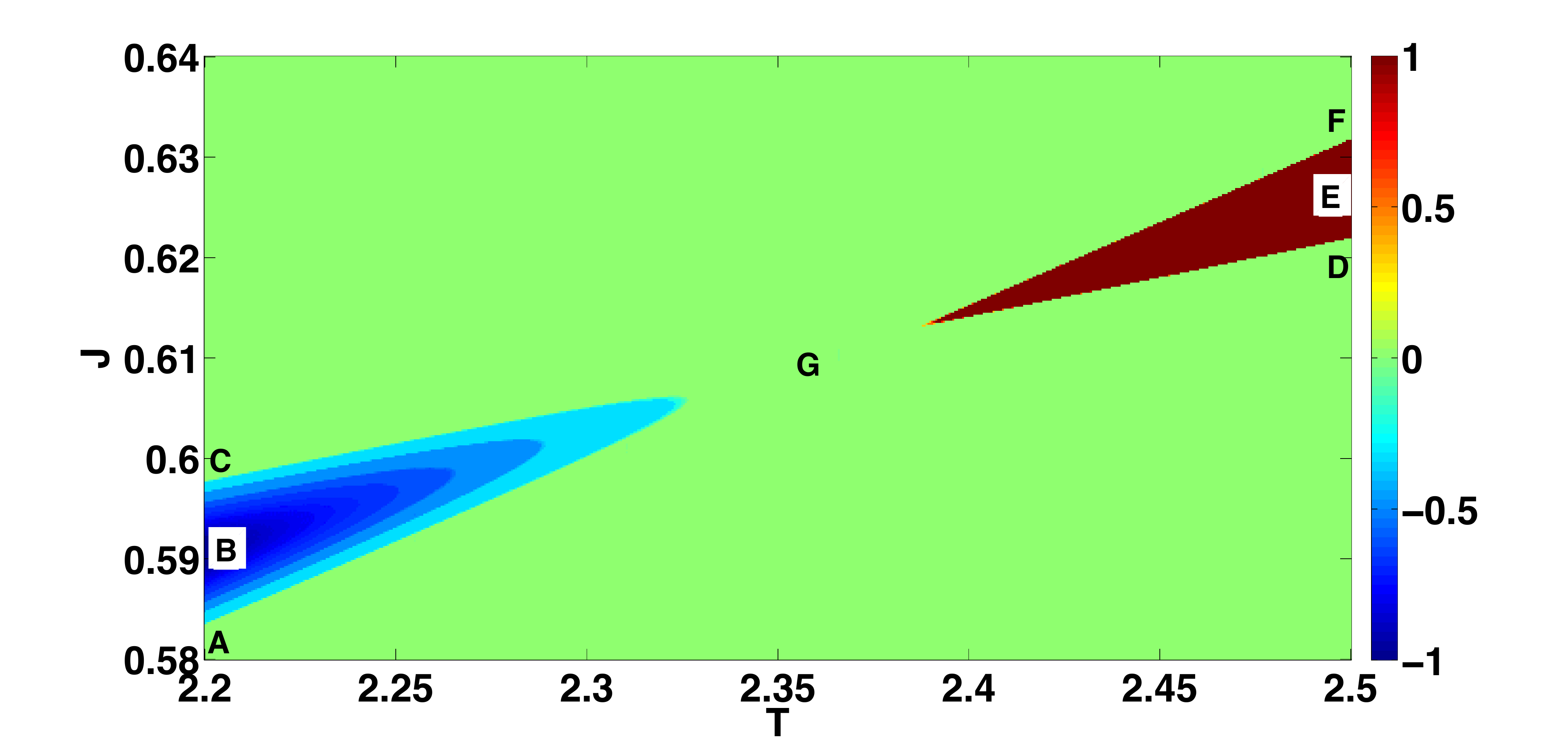}
\end{tabular}
\caption{Detail of the bifurcation regions of $\langle v\rangle $ in the parameter's space $[J,T]$ for positive and negative delta trains of length $K=130$. The significance of points $A$ to $G$ are explained in the text}.
\label{fig:inversion}
\end{figure}

Let us analyze Fig. \ref{fig:inversion}. This is a zoom in the current reversal area of the mean velocity bifurcation pattern shown in Fig. \ref{fig:avvel}d

Let us consider seven points, $A$ through $G$:
\begin{itemize}
\item Point $A$: $T=2.22$, $J=0.580$, $<v>/v_{\omega}=0$ 
\item Point $B$: $T=2.22$, $J=0.590$, $<v>/v_{\omega}=-1 $
\item Point $C$: $T=2.22$, $J=0.600$, $<v>/v_{\omega}=0 $
\item Point $D$: $T=2.50$, $J=0.617$, $<v>/v_{\omega}=0 $
\item Point $E$: $T=2.50$, $J=0.627$, $<v>/v_{\omega}=+1 $
\item Point $F$: $T=2.50$, $J=0.637$, $<v>/v_{\omega}=0 $
\item Point $G$: $T=2.36$, $J=0.610$, $<v>/v_{\omega}=0$
\end{itemize}

For points $A$ and $D$: when the $K=130$ positive Dirac deltas are applied, the normalized impulse value $J/(2\pi\gamma) $ is not high enough to carry particles from around the $\tilde{x}=0.1904$ stable fixed point to the right of the $\tilde{x}=0.8096$ unstable fixed point (see Fig. \ref{fig:fijosYsaltos}). Then, the force-positive-half-cycle idle time begins, and the particle returns to a position near the starting point $\tilde{x}=0.1904$. Now the $K=130$ negative Dirac deltas are applied, and the impulse value $J$ is not high enough to carry particles from around the $\tilde{x}=0.1904$ stable fixed point to the left of the $\tilde{x}=-0.1904$ unstable fixed point. Then, the force-negative-half-cycle idle time begins, and the particle returns again to around the $\tilde{x}=0.1904$ starting fixed point. Then, the average velocity $<v>/v_{\omega}$ equals zero.

For points $C$ and $F$: when the $K=130$ positive Dirac deltas are applied, the normalized impulse value $J/(2\pi\gamma) $ is high enough to carry particles from around the $\tilde{x}=0.1904$ fixed point to the right of the $\tilde{x}=0.8096$ unstable fixed point (see Fig. \ref{fig:fijosYsaltos}). Then, the force-positive-half-cycle idle time begins, and the particle goes forward to around the $\tilde{x}=1.1904$ starting fixed point. Now the $K=130$ negative Dirac deltas are applied, and the impulse value $J/(2\pi\gamma) $ is high enough to carry particles from around the $\tilde{x}=1.1904$ fixed point to the left of the $\tilde{x}=0.8096$ unstable fixed point. Then, the force-negative-half-cycle idle time begins, and the particle goes back around the $\tilde{x}=0.1904$ starting fixed point. Then, the average velocity equals zero.

The question now is: why does point $B$ have a negative velocity and why does point $E$ have a positive velocity. In order to understand that, we have to take into account that a \textsl{relaxation time} $t_r=T/1000$ follows every Dirac delta we apply. Let us stress that $t_r$ is the small time between two consecutive Dirac deltas, and not the idle time $t_{iddle}$ mentioned above.  Since the  absolute value of the ratchet force is bigger in the $0.8096< \tilde{x}<1.1904$ region than in the $0.1904< \tilde{x}<0.8096$ region, the ratchet force has more influence during the $K-1$ relaxation times in the negative half cycle than during the positive half cycle. An increase in $J/(2\pi\gamma) $ from point $D$ to point $E$ enables particles  to go from around the $\tilde{x}=0.1904$ fixed point to the right of the $\tilde{x}=0.8096$ unstable fixed point when the positive Dirac deltas are applied. Then, the positive-half-cycle idle time begins, and particles move forward  to the $\tilde{x}=1.1904$ stable fixed point. When the negative Dirac deltas are applied, however, the influence of the ratchet force is so important, that the particles are not able to go from $\tilde{x}=1.1904$ fixed point and reach the left of the $\tilde{x}=0.8096$ unstable fixed point. Then, the negative-half-cycle idle time begins, and particles return around the $\tilde{x}=1.1904$ fixed point. Therefore, the average velocity is positive.

Let us consider now points $A$, $B$, and $C$ 
Their periods are lower than the periods of points, $D$, $E$, and $F$. The lower the period, the lower the relaxation time, and the lower the influence of the ratchet force on the particle behavior.
An increase in $J/(2\pi\gamma) $ from  point $A$ to point $B$ is not enough to make the particle go from around the $\tilde{x}=0.1904$ stable fixed point to the right of the $\tilde{x}=0.8096$ unstable fixed point, when the positive Dirac deltas are applied. Then, the positive-half-cycle idle time begins, and particles return around the $\tilde{x}=0.1904$ starting fixed point. When the negative Dirac deltas are applied, however, the influence of the ratchet force during $t_r$ is not so important and the particles are able to go from around the $\tilde{x}=0.1904$ fixed point to the left of the $\tilde{x}=-0.1904$ unstable fixed point. Then, the negative-half-cycle idle time begins, and particles go around the $\tilde{x}=-0.8096$ stable fixed point. Therefore, the average velocity is negative.

Now consider point $G$ . For those $J/(2\pi\gamma) $ and $T$ values, the Dirac deltas are just large enough to carry particles from the $\tilde{x}=0.1904$ stable fixed point to the $\tilde{x}=0.8096$ unstable fixed point during the application of the force-positive-half-cycle Dirac deltas. They also carry particles from the $\tilde{x}=1.1904$ stable fixed point to the $\tilde{x}=0.8096$ unstable fixed point during the application of the force-negative-half-cycle Dirac deltas.

Finally,  let us stress that if $K$ is increased to $500$, the $\alpha=1$  behavior is almost identical to that obtained with a square wave with amplitude $\pm A$ and period $T$,  with $AT/2=J=KJ_i$.

\begin{figure}
\begin{tabular}{cc}
\includegraphics[width=1\textwidth]{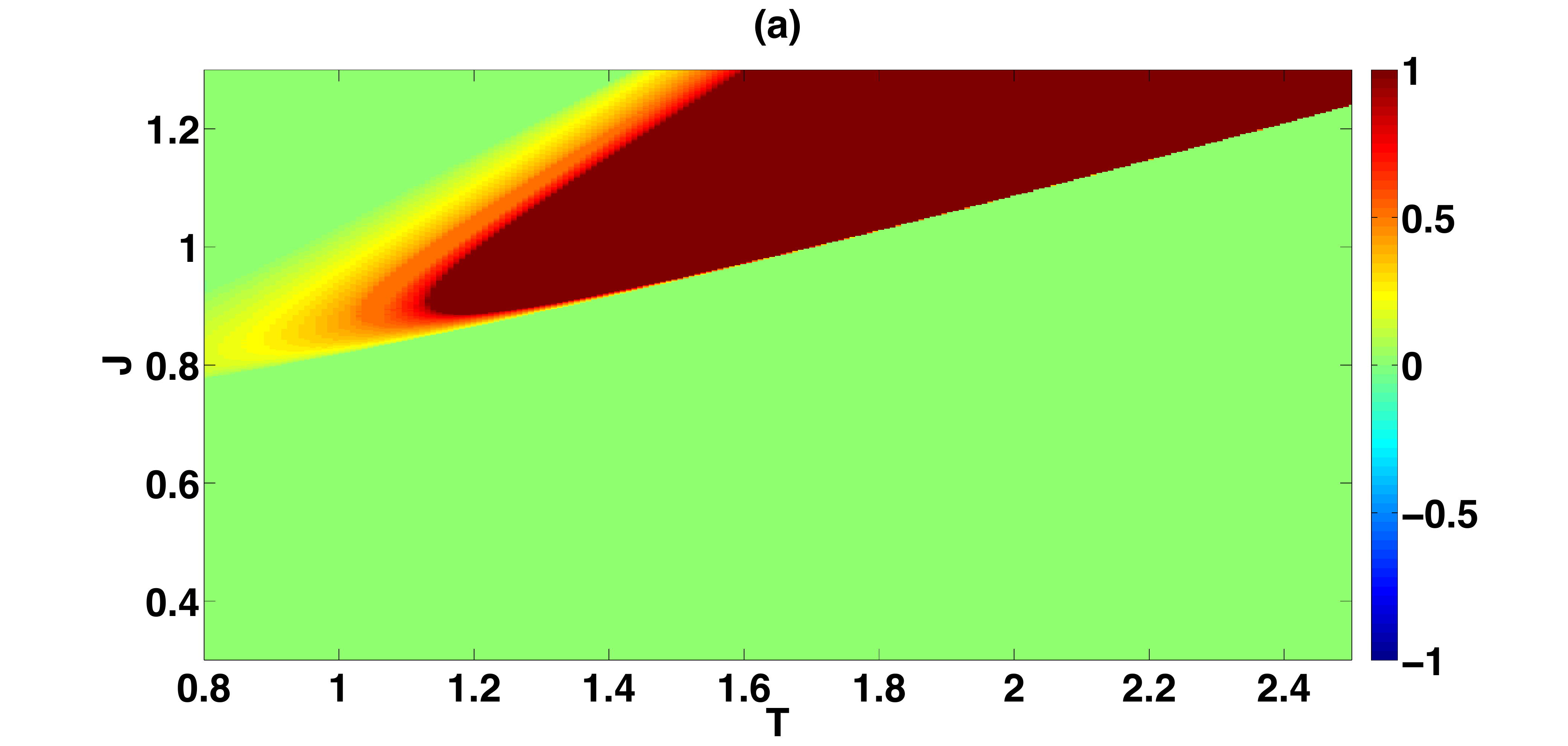}\\
\includegraphics[width=1\textwidth]{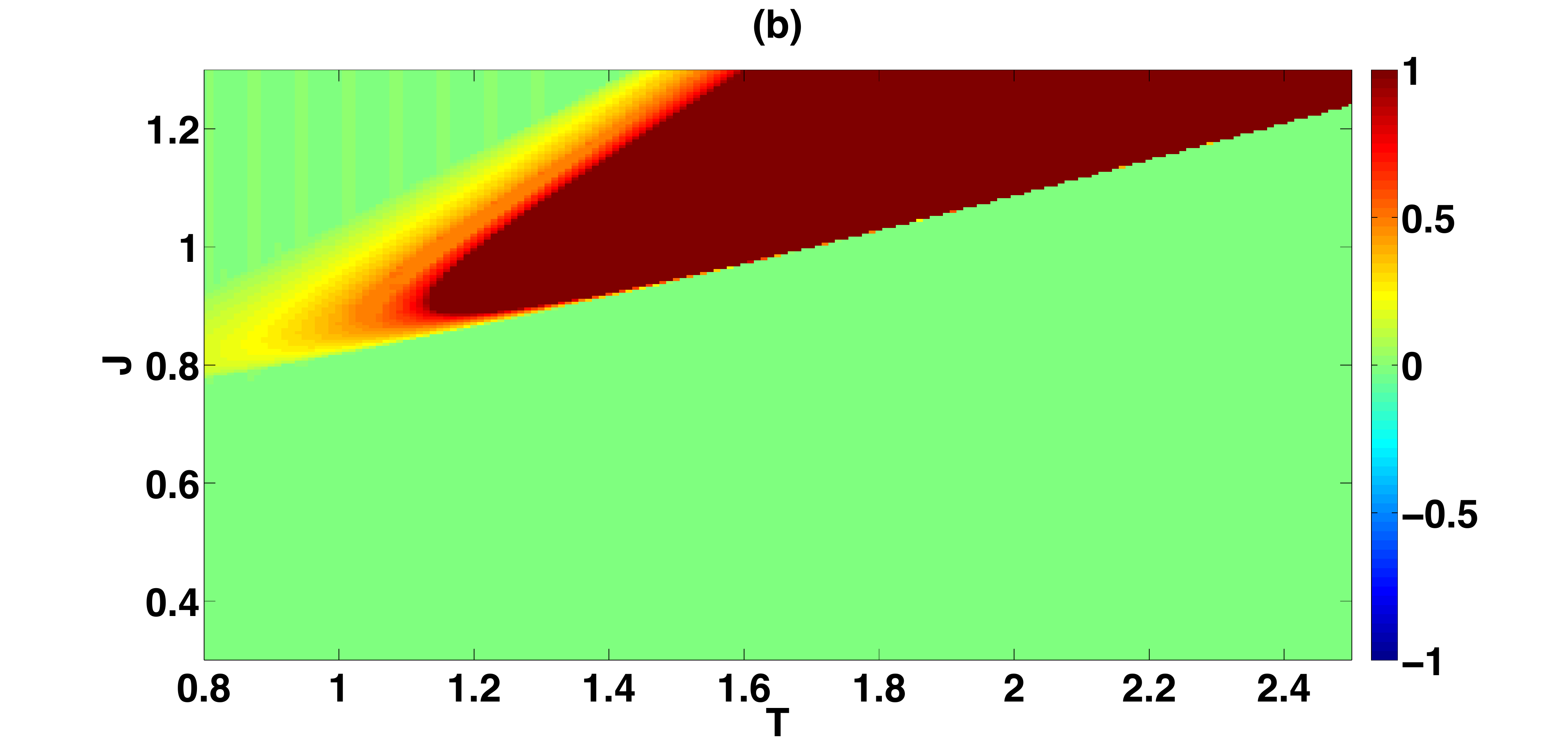}\\
\end{tabular}
\caption{Bifurcation regions of the mean velocity in the parameter space $[J,T]$. The system is the  overdamped ratchet with $\gamma=0.1109$ and $\mu=0.5$; (a) $500$ consecutive positive deltas at $t=0+iT/1000+nT, i=1,..., 499, n=1,...100$  and $500$ consecutive negative deltas at $t=T/2+iT/1000+nT, i=1,..., 499, n=1,...,100$; (b) square waveform with a positive semicycle of length $T/2$ and amplitude $A=1000 J/T$, and a negative semi cycle with amplitude $-A$}.
\label{fig:deltavscuadrada}
\end{figure}

Figs. \ref{fig:deltavscuadrada}(a) and  \ref{fig:deltavscuadrada}(b) show the bifurcation diagram for the mean velocity in the $J-T$ parameter space for 500 delta function pulses and for a square wave. It is clear from these results that the behavior of the system is almost identical for both cases.

\section{Conclusions}
\label{sec:conclusions}

We have introduced a {\it minimal} one-dimensional deterministic ratchet model
in the overdamped regime driven by alternate positive and negative pulses.
The strong nonlinearity of the driving force produces a bifurcation pattern
with synchronized as well as chaotic regions. The integrability of delta functions
allowed us to obtain analytical maps of the particle dynamics and study the
transition from regular to chaotic motion. We find that a tangent bifurcation is
associated with this transition.  Both our analytical 1-D map and the power-law behavior of the corresponding
power spectrum confirm this. In order to extend the  analysis to other more general types of periodic forces we consider a set of $K$ alternate positive and $K$ negative pulses as the driving force. A transition from negative to positive current is obtained with increasing $K$. Finally, we find that the synchronization regions of
both a continuous square wave and a pulse composed of a series of delta function pulses have
equivalent dynamical behavior, indicating that the continuous driving force may be considered as
a succession of pulses from the point of view of the ratchet dynamical system
we have studied.

\textbf{Acknowledgements}
This work was partially supported by CONICET, ANPCyT and Universidad Nacional de Mar del Plata.

%
\bibliographystyle{apsrev4-1}
\bibliography{HALbibratchet}

%
%
%
%
%
%
%
%
%

\end{document}